\documentclass[11pt,a4paper]{article}
\usepackage{jheppub}
\usepackage{slashed}
\usepackage{amsmath}
\usepackage{amssymb}
\usepackage{epsfig}
\usepackage{mathrsfs}

\def\({\left(} 
\def\){\right)}
\def\[{\left[} 
\def\]{\right]}
\newcommand{\non}{\nonumber \\}

\newcommand{\ie}{{\it i.e.,}\ }
\newcommand{\eg}{{\it e.g.,}\ }

\newcommand{\Z}{{\mathbb Z}}

\newcommand{\be}{\begin{equation}}
\newcommand{\ee}{\end{equation}}
\newcommand{\bea}{\begin{eqnarray}}
\newcommand{\eea}{\end{eqnarray}}

\newcommand{\mt}[1]{\textrm{\tiny #1}}

\newcommand{\bt}{\beta}

\catcode`\@=12

\newcommand{\al}{\alpha}

\def\del          {\partial}

\newcommand{\ad}[1]{\textcolor{blue}{#1}}

\newcommand{\reef}[1]{(\ref{#1})}
\renewcommand{\eqref}[1]{(\ref{#1})}

\def\ph1{\phantom{1}}

\newcommand{\beq}{\begin{equation}}
\newcommand{\eeq}{\end{equation}}
\newcommand{\ba}{\begin{aligned}}
\newcommand{\ea}{\end{aligned}}
\newcommand{\beqa}{\begin{eqnarray}}
\newcommand{\eeqa}{\end{eqnarray}}
\newcommand{\beqar}{\begin{eqnarray*}}
\newcommand{\eeqar}{\end{eqnarray*}}

\begin{document}

\title{A model of persistent breaking of continuous symmetry}


\author[1]{Noam Chai,}
\author[2,3]{Anatoly Dymarsky,}
\author[1,4]{Mikhail Goykhman,}
\author[1]{Ritam Sinha,}
\author[1]{and Michael Smolkin}

\affiliation[1]{The Racah Institute of Physics, The Hebrew University of Jerusalem, \\ Jerusalem 91904, Israel \\}
\affiliation[2]{Department of Physics and Astronomy, \\ University of Kentucky, Lexington, KY 40506\\[2pt]}
\affiliation[3]{Skolkovo Institute of Science and Technology, \\ Skolkovo Innovation Center, Moscow, Russia, 143026\\[2pt]}
\affiliation[4]{William I. Fine Theoretical Physics Institute, University of Minnesota, Minneapolis, MN
55455, USA}

\emailAdd{noam.chai@mail.huji.ac.il}
\emailAdd{a.dymarsky@uky.edu}
\emailAdd{goykhman@umn.edu}
\emailAdd{ritam.sinha@mail.huji.ac.il}
\emailAdd{michael.smolkin@mail.huji.ac.il}

\abstract{We 
consider a UV-complete field-theoretic model in general dimensions, including $d=2+1$, which consists  of two  copies of the
long-range vector models, with $O(m)$ and $O(N-m)$ global symmetry groups,
perturbed by double-trace operators.
Using conformal perturbation theory
we find weakly-coupled IR fixed points for $N\geq 6$ that reveal a spontaneous
breaking of global symmetry. Namely, at finite temperature the lower rank group is broken,
with the pattern persisting at all temperatures due to scale-invariance. 
We provide evidence that the models in question are unitary and  invariant under  full conformal symmetry. 
Furthermore, we show that this model exhibits a continuous family of weakly interacting field theories at finite $N$.
}

\maketitle

\section{Introduction}

The phenomenon of persistent symmetry breaking (PSB), \textit{i.e.}, spontaneous breaking of a global symmetry that persists at arbitrarily high temperatures, was first noticed by Weinberg \cite{Weinberg:1974hy}. Since then it has been actively discussed in cosmology \cite{Dvali:1995cj,Dvali:1995cc,Senjanovic:1998xc, Bajc:1999cn, Ramazanov:2021eya,Ramazanov:2020ajq},  quantum field theory \cite{Hong:2000rk,Komargodski:2017dmc,Chai:2020onq,Chai:2020zgq,Chai:2020hnu,Chai:2021djc,Chaudhuri:2020xxb,Bajc:2020gpa,Chaudhuri:2021dsq} and holography \cite{Buchel:2009ge,Donos:2011ut,Gursoy:2018umf,Buchel:2018bzp,Buchel:2020thm,Buchel:2020xdk,Buchel:2020jfs,Buchel:2021ead}. 

The AdS/CFT candidates for persistent order are perturbatively stable, but the symmetric phase has smaller free energy.  In fact, there are a number of theoretical results, which  guarantee symmetry restoration at sufficiently high temperatures  in certain particular settings, yet in 
some models the PSB behavior is possible. Thus, during the last year a number of papers studied critical points of the bi-conical
$O(m)\times O(N-m)$  model
in the context of the PSB.
At finite temperature $T$, these models exhibit 
spontaneous symmetry breaking, and because of scale-invariance symmetry breaking  persists to an arbitrarily high $T$  \cite{Chai:2020onq,Chai:2020zgq,Chai:2020hnu,Chai:2021djc}. Furthermore, the authors of \cite{Chaudhuri:2020xxb,Chaudhuri:2021dsq} constructed critical gauge theory models in $3+1$ dimensions which exhibit symmetry breaking  at arbitrary high temperatures in the infinite $N$ limit. 
Note that starting from a scale invariant model resolves the issue of UV completeness,  present in the original example of  PSB  \cite{Weinberg:1974hy}. 
There are also examples of non-unitary models of persistent breaking in the presence of chemical potential \cite{Hong:2000rk,Komargodski:2017dmc,Tanizaki:2017qhf,Dunne:2018hog,Wan:2019oax}.\footnote{Persistent order can  be also considered in the context of spontaneous breaking of higher-form symmetries, see {\it e.g.,} \cite{Aitken:2017ayq}. In this work we focus on the ordinary (zero-form) symmetries.}

The aforementioned works suggest that persistent breaking of global symmetries is possible. Therefore, the standard logic which suggests that the disordered phase has larger entropy ${\cal S}$, and therefore for sufficiently large $T$ the disordered phase would have  smallest free  energy
${\cal F} = {\cal E}- T \,{\cal S}$, under certain circumstances can break down.

In this paper, we study persistent symmetry breaking
in critical models with the long-range interactions. While the models of interest can  be defined in general dimensions $1<d<4$ ($d$ is bounded to ensure the stability of the model, fixed by the unitarity bound), our primary interest
will be in physically motivated case of $d=2+1$.
In $2+1$ dimensions, the Coleman-Hohenberg-Mermin-Wagner (CHMW) theorem
prohibits spontaneous breaking of continuous symmetries at non-zero temperature \cite{Mermin:1966fe,Hohenberg:1967zz,Coleman:1973ci}. 
However, the necessary assumptions for the CHMW result can  be evaded,
for instance, by focusing on spontaneous breaking of discrete symmetries.
This is what was done in \cite{Chai:2021djc}, which studied  persistent breaking of the global  $\mathbb{Z}_2$ symmetry
in the $\mathbb{Z}_2\times O(N)$ long-range vector model.
Moreover, the continuous
symmetry in similar models can also be broken persistently in $d=2+1$,
owing to yet another way around the assumptions behind the CHMW theorem:
by considering systems with long-range interactions \cite{Hohenberg:1967zz,Mermin:1968zz,Halperin_2018}.\footnote{This
is to be contrasted with the earlier models of PSB \cite{Chai:2020onq,Chai:2020zgq,Chai:2020hnu},
that studied the local bi-conical $O(m)\times O(N-m)$ vector models, where
continuous symmetry breaking can occur only in non-integer dimensions $3<d<4$. The latter
choice of dimension, however, raises the issue of unitarity \cite{Hogervorst:2015akt}.}
In this paper the bi-conical long-range critical $O(m)\times O(N-m)$ vector model is studied
perturbartively, close to the Gaussian mean field theory point. We  explore IR  flows and properties of the IR fixed points at zero temperature. 
As a result interesting in its own right, we show that our model in $3d$ exhibits a conformal manifold of interacting fixed points at infinite $N$.  At finite $N$ the continuous space of interacting fixed points persists and is parametrized by a continuous family of Gaussian theories in the UV. 

One of the main advantages of considering long-range interactions
is the possibility to formulate a model admitting  perturbative treatment in arbitrary number of dimensions $d$. This is achieved by the choice of the critical exponent of the bi-local kinetic term of the generalized free field. Subsequently turning on a quartic coupling results in an RG flow, terminating at a weakly-coupled IR fixed point, located in the perturbative vicinity of the long-range mean field theory point \cite{Fisher:1972zz}. While this approach bears certain similarity to the Wilson-Fisher $\epsilon$-expansion \cite{Wilson:1971dc},
it is applicable in any $d$, including $d=3$.

An important feature of the long-range models is absence of the local stress  tensor.
Without the stress  tensor full conformal symmetry of the scale-invariant fixed point is not manifest. An extensive supporting argument
in favor of the full conformal invariance of the RG fixed point of the long-range Ising
model (i.e., $O(N)$ vector model with $N=1$) has recently been 
put forward  in  \cite{Paulos:2015jfa,Behan:2017dwr,Behan:2017emf,Behan:2018hfx}.
For large $N$, the long-range $O(N)$ vector model at criticality has been studied
in \cite{Gubser:2017vgc,Giombi:2019enr} using $1/N$ expansion, with the most recent works 
\cite{Chai:2021arp,Chakraborty:2021lwl} providing a strong evidence that the critical
regime of this model is in fact described by a CFT (see also
\cite{Brydges:2002wq,Abdesselam:2006qg,Brezin2014,Slade:2016yer,Gubser:2017vgc,Giombi:2019enr,Benedetti:2020rrq}
for previous calculations of critical exponents).\footnote{
It was argued in \cite{Behan:2017dwr,Behan:2017emf} that the non-local
CFT describing the IR fixed point of the long-range $\phi^4$ model
can also be found at the IR end-point of an RG flow triggered by coupling
the short-range vector field to the generalized free field of dimension $(d+s)/2$.
Such a coupling is irrelevant in the short-range regime,
and can be studied perturbatively near the long-range to short-range
crossover point.
 $O(N)$
generalization of this IR duality has recently been discussed in \cite{Chai:2021arp}.
}
At the same time, full conformal invariance  of the critical bi-conical long-range $O(m)\times O(N-m)$ model is less understood. 
Full conformal symmetry, if present, would restrict the functional form
of the three-point functions of primary operators, and ensure that cross-correlators of primaries
with different conformal dimensions vanish. We 
perform several consistency checks of the long-range $O(m)\times O(N-m)$ vector models at criticality and 
 confirm expected behavior of the  two and three-point functions, dictated by full conformal symmetry. In the process, we verify that the anomalous dimensions of all considered single-trace and double-trace
  operators remain real, which is a necessary condition of unitarity. 

This  paper is organized as follows.
In section~\ref{sec:model} we begin by defining our model.
Working at the linear order in perturbation theory near free critical point, we derive the RG flow equations
for the quartic double-trace interaction coupling constants. Then
we analyze the fixed points of the RG flow for different choices
of scaling dimensions of the scalar fields vector multiplets $\phi_{1,2}$. In particular, we
focus on the models admitting negative fixed point value of the coupling constant $g_3$
corresponding to the quartic operator $\phi_1^2\phi_2^2$. Behavior of such
fixed points at finite temperature is then explored in section~\ref{sec: thermal physics}. Specifically,
we demonstrate that some of the fixed points $g_3<0$ lead to an instability of the
symmetric vacuum $\phi_1=0$ of the effective action at $T>0$. The resulting model
therefore breaks $O(m)$ symmetry spontaneously at any non-zero temperature,
exhibiting the phenomenon of PSB.

The model in question admits a continuous family of interacting fixed points, that we discuss in section~\ref{sec:confmanifold}. 
We additionally 
explore the nature of the critical regime
of the considered bi-conical model at zero temperature in section~\ref{sec:anomalous dimensions},
where we calculate anomalous dimensions of various single-trace (quadratic) and double-trace (quartic) operators.
To this end, we take into account the operator mixing effect, and diagonalize the
correlation matrices by finding the true basis of primary operators. As we mentioned above,
fixed point of an interacting long-range model, lacking a local stress-energy tensor, 
might end up being scale-invariant but not conformal invariant. 
We carry out several checks of the full conformal symmetry in section~\ref{sec:tests of conformal invariance}.
In that section, we calculate cross-correlator between quadratic and quartic conformal
primaries, and demonstrate that they vanish at the considered order in $\epsilon$-expansion.
We also calculate three-point correlator and show that  at leading order it agrees with
the form dictated by the conformal symmetry.  

The results of the paper are summarized in section~\ref{sec:discussion}.

\section{Long range $O(m)\times O(N-m)$ vector model at criticality}
\subsection{The model and the RG flow}
\label{sec:model}

Consider the following Gaussian action in $1\leq d < 4$ dimensions \footnote{The range of $d$ ensures the stability of of the model, and is fixed by the unitarity bound, $d/4>\Delta_\phi \geq \frac{d-2}{2}$},
\be
\label{S0}
 S_0= \mathcal{N}_1 \int d^d x_1 \int d^d x_2 {\vec \phi_1(x_1)\cdot \vec\phi_1(x_2) \over |x_1-x_2|^{2(d-\Delta_{\phi_1})}}
 +\mathcal{N}_2 \int d^d x_1 \int d^d x_2 {\vec\phi_2(x_1)\cdot \vec\phi_2(x_2) \over |x_1-x_2|^{2(d-\Delta_{\phi_2})}} ~.
\ee
The model (\ref{S0}) describes two real-valued generalized free scalar fields $\vec \phi_1$ and $\vec \phi_2$ transforming in  vector representation of the $O(m)$ and $O(N-m)$ global symmetry groups.
Our conventions are such that $m < N$.
The coefficients $\mathcal{N}_{1,2}$
are fixed so that the two point functions 
of $\vec \phi_{1,2}$ in position space are normalized to one.  The scaling dimensions of the generalized free fields  are
\be
 \Delta_{\phi_i}={d-\epsilon_i\over 4}\,,  \quad  i=1,2~.
\ee
For brevity, we will be suppressing $O(m)$, $O(N-m)$ vector indices below.

In what follows, we are going to consider deformation of the free action (\ref{S0})
by the following  double-trace operators
\be
 \mathcal{O}_1= (\phi_1^2)^2 ~, \mathcal{O}_2= (\phi_2^2)^2, ~ \mathcal{O}_3= \phi_1^2 \phi_2^2~.
\ee
Choosing $\epsilon_i\ll 1$, one can make these operators weakly relevant, 
with the leading order scaling dimensions $\Delta_1=4\Delta_{\phi_1}$, $\Delta_2=4\Delta_{\phi_2}$ and $\Delta_3=2(\Delta_{\phi_1}+\Delta_{\phi_2})$.  At the same time,
the leading-order two-point functions of these operators are given by
\bea
 && \langle \mathcal{O}_i(x) \mathcal{O}_j(0)\rangle = \delta_{ij}\,  {N_i\over |x|^{2\Delta_i}},
 \non
&& N_1 = 8 m^2\Big(1+{2\over m}\Big),\, N_2 = 8 (N-m)^2\Big(1+{2\over N-m}\Big),\, N_3= 4m(N-m)~.
 \label{2p}
\eea
Similarly, the leading-order three-point functions 
\bea
\nonumber
\langle\mathcal{O}_i(x_1) \mathcal{O}_j(x_2) \mathcal{O}_k(x_3)\rangle&=&{C_{ij}^k N_k \over
|x_{12}|^{\Delta-2\Delta_k} 
|x_{23}|^{\Delta-2\Delta_i} 
|x_{13}|^{\Delta-2\Delta_j}},\\
\Delta&=&\Delta_i+\Delta_j+\Delta_k,
\eea
are fixed by the OPE coefficients (we list only the non-zero ones)
\begin{equation}
\label{OPE coefficients}
\begin{split}
 &C_{11}^1= 8 \,(m+8) ~, ~ C_{33}^1=2(N-m)~, ~ C^3_{13}=4\(m+2\)~,~ C^3_{33}=16~,\\
 &C^3_{32}=4(N-m+2)~, ~ C_{33}^2=2m ~, ~ C_{22}^2=8 \, (N-m+8) ~.
\end{split}
\end{equation}
The latter are related by
\be
C_{ij}^k =C_{ik}^j N_j/N_k.
\label{OPErelation}
\ee

\noindent
Consider now the following deformation of the Gaussian theory (\ref{S0})
\be
 S=S_0 + \sum_{i=1}^3 {g_i \mu^{\epsilon_i} \over N} \int d^dx \, \mathcal{O}_i(x)  ~,
 \label{Sdeform}
\ee
where $\mu$ is an arbitrary RG scale, and we also denoted $\epsilon_3=(\epsilon_1+\epsilon_2)/2$.
This deformation induces an RG flow, that at the next-to-leading (one-loop) order in perturbation
around the free regime has the form
\be
 \mu {dg_i\over d\mu} = -\epsilon_i g_i + {\pi^{d/2} \over N\Gamma\({d\over 2}\)} \sum_{j,k} C_{jk}^i g_j g_k + {\cal O}(g_i^3)\,.
 \label{RGflow}
\ee
We are interested in the interacting IR critical regime of the model (\ref{Sdeform}).
To this end, we need to determine  fixed-points of the RG flow (\ref{RGflow}).
At the one-loop order, the critical parameters can be found by plugging in the values of the OPE coefficients
(\ref{OPE coefficients}) into the r.h.s. of (\ref{RGflow}). One of the fixed points, with $g_3=0$, describes two decoupled copies of the
 long-range vector models.
 We will not be considering this fixed point it in what follows.
When $g_3\neq 0$ the fixed points can be found by solving the following system of coupled second-order equations,
\begin{equation}
\begin{aligned}
\tilde g_1&= {C_{11}^1\over N} ~ \tilde g_1^2 + {C_{33}^1\over N} \Big({\alpha+1\over 2\alpha}\Big)^2 ~ \tilde g_3^2 ~,\\
 \tilde g_2&= {C_{22}^2\over N} ~ \tilde g_2^2 + {C_{33}^2\over N} \Big({\alpha+1\over 2}\Big)^2 ~ \tilde g_3^2   ~,\\
1&= {C^3_{13}\over N} \,{4\alpha\over \alpha+1} \, \tilde g_1 + {C^3_{33}\over N} \tilde g_3 + {C^3_{32}\over N} \, {4\over \alpha+1} \,  \tilde g_2~.
\end{aligned}
\label{beta-fn}
\end{equation}
Here we have rescaled the couplings as $g_i = \tilde g_i {\Gamma\({d\over 2} \) \over \pi^{d/2}} \epsilon_i$, and defined the
parameter
\begin{equation}
\alpha = \frac{\epsilon_1}{\epsilon_2}\,.
\end{equation}

\noindent 
Notice that the equations (\ref{beta-fn})  are invariant under
the redefinitions
$\al\to1/\al, \, m\to N-m$, $\tilde g_1 \leftrightarrow \tilde g_2$,
originating from a simple interchange of notations for the fields $\phi_{1,2}$
in the original action (\ref{Sdeform}).
Therefore there is a one-to-one correspondence between the family
of solutions with $\al\geq 1$ and $\al\leq 1$. Without loss of generality we restrict our analysis to the case $\al\geq 1$. Another thing to point out is that $\alpha$ is a free parameter, just like $N$ and $m$ and is fixed by construction. Since the scalar fields do not acquire anomalous dimensions, $\alpha$ is unaffected by the dynamics of the model.

It can be easily seen, as we discuss below in section \ref{sec: thermal physics}, that the persistent symmetry breaking is only possible when the fixed-point value of the coupling constant $g_3$ is negative. These are the solutions of \eqref{beta-fn} that we wish to explore.

\begin{figure}
\includegraphics[width=0.95\textwidth]{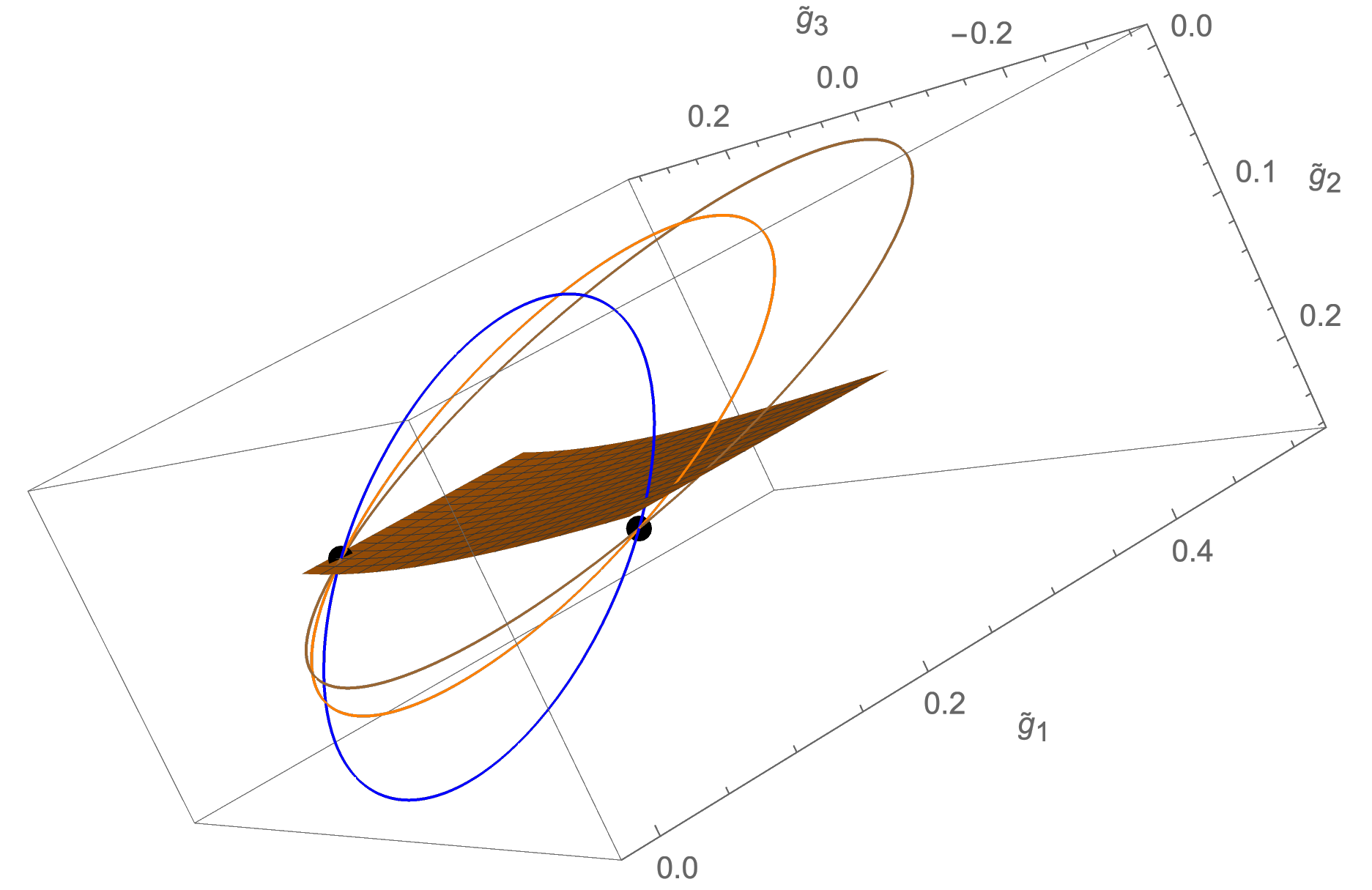}
\caption{Conformal manifolds \eqref{branches} for $x_1=1/2$ (blue), $x_1=1/3$ (orange) and $x_1=1/4$ (brown). Black points correspond to 
$\tilde g_1=\tilde g_2=1/8, \, \tilde g_3=\pm 1/4$ which are common for all $x_1$. Orange surface, \eqref{surface} with $x_1=1/3$, intersects the orange curve at two points (with positive and negative $\tilde g_3$) which are two  theories with  $m/N=1/3$, $\alpha=1$, and large but finite $N$. Theory with positive $ \tilde g_3 = 1/4$ has the global symmetry enhanced to $O(N)$, while the theory with negative $\tilde g_3 = -1/4$ at finite temperatures exhibits persistent symmetry breaking $O(m)\times O(N-m)\rightarrow O(m-1)\times O(N-m)$.}
\label{fig:infiniteN}
\end{figure}

While in general it is difficult to solve the system of second-order equations (\ref{beta-fn})
analytically, a simplification can be achieved in the large $N$ limit. We begin by considering
the case of $N\gg 1$, $m\gg 1$, with fixed $m/N = {\cal O}(1)$. Denoting
\begin{equation}
x_1 = \frac{m}{N}\,,\qquad x_2 = 1-x_1\,,
\end{equation}
we can re-write equations (\ref{beta-fn}) at the leading order in $1/N$ as
\begin{equation}
\begin{aligned}
\tilde g_1&= 8 x_1 ~ \tilde g_1^2 +  2 x_2 \Big({\alpha+1\over 2\alpha}\Big)^2 ~ \tilde g_3^2 ~,\\
 \tilde g_2&= 8 x_2 ~ \tilde g_2^2 + 2 x_1  \Big({\alpha+1\over 2}\Big)^2 ~ \tilde g_3^2   ~,\\
1 &= {16  \over \al + 1} \(  \al \, x_1  \, \tilde g_1  +x_2 \,  \tilde g_2 \)~.
\end{aligned}
\label{RGinfN}
\end{equation}
These equations admit a solution only if $\alpha=1$, in which case they are degenerate, and
a line of fixed points consisting of two branches emerges,\footnote{
This solution was in fact first found in \cite{Chai:2020hnu}, that considered the bi-conical
$O(m)\times O(N-m)$ model in $4-\epsilon$ dimensions. At the one-loop
order in perturbation theory, the fixed point of this model is determined by the
equations (\ref{beta-fn}) with $\alpha=1$.}
\bea
 &&\tilde g_1^{\pm}= {1\pm\sqrt{1-64 x_1 x_2 \tilde g_3^2}\over 16 x_1} ~,
 \non
  &&\tilde g_2^{\pm}= {1\mp\sqrt{1-64 x_1 x_2 \tilde g_3^2}\over 16 x_2}~,
  \label{branches}
  \\
  &&\tilde g_3\in \Big[-{1\over 8\sqrt{x_1x_2}}~, ~ +{1\over 8\sqrt{x_1x_2}}\Big].
  \nonumber
\eea 
The two branches meet at the end points $\tilde g_3= \pm 1/(8\sqrt{x_1x_2})$, forming a circle, see Fig.~\ref{fig:infiniteN}. The coordinate on this circle parametrizes one of the directions
on the two-dimensional conformal manifold of the model (\ref{Sdeform}); we will discuss
the latter in more
detail in section~\ref{sec:confmanifold}.
The corresponding exactly marginal
operator ${\cal O}_+'$, obtained by mixing the double-trace operators ${\cal O}_i$, $i=1,2,3$,
will be derived below in section~\ref{sec:anomalous dimensions}. 

There is a special point  $\tilde g_1=\tilde g_2=1/8, \, \tilde g_3=1/4$  where $O(m)\times O(N-m)$ symmetry is enhanced to 
the full  $O(N)$. Being independent of $x_1$, it can be found as the intersection point of curves with different $x_1$. 
Since the RG equations in the infinite $N$ limit, given by \eqref{RGinfN}, are invariant
w.r.t. $\tilde g_3\rightarrow -\tilde g_3$, another intersection point is given by
$\tilde g_1=\tilde g_2=1/8, \, \tilde g_3=-1/4$.

Additionally, each curve has 
 ``decoupled'' points $\tilde g_1^-=\tilde g_3^-=0, \tilde g_2^-={1\over 8x_2}$ and $\tilde g_2^+=\tilde g_3^+=0, \tilde g_1^+={1\over 8x_1}$ where only one critical long-range vector model survives.

To find corrections to the large-$N$ solution \eqref{branches}
at the next-to-leading order in $1/N$ expansion,
we substitute the ansatz
\begin{equation}
\tilde g_i \to \tilde g_{i} + \delta\tilde g_{i}/N + \mathcal{O}(1/N^2)\,,\quad 
\alpha=1 + \delta\alpha/N +  \mathcal{O}(1/N^2)
\end{equation}
into (\ref{beta-fn}), and linearize over the $1/N$ terms.
This yields
\begin{equation}
\label{subleading 1/N g's}
 \left(\begin{array}{ccc} 1-16 \tilde g_1 x_1 & 0 & -4\tilde g_3 x_2 \\
 0 & 1-16 \tilde g_2 x_2 & -4 \tilde g_3 x_1  \\ 2x_1 & 2x_2 & 0\end{array}\right)
 \left(\begin{array}{c} \delta \tilde g_1 \\ \delta \tilde g_2 \\ \delta \tilde g_3 \end{array}\right)
 =
 \left(\begin{array}{c} 64\tilde g_1^2  \\ 64 \tilde g_2^2 \\ - 4\sum_{i=1}^3 \tilde g_i \end{array}\right) 
 +
 \left(\begin{array}{c} -2 \tilde g_3^2 x_2  \\ 2 \tilde g_3^2 x_1 \\ \tilde g_2-(\tilde g_1+\tilde g_2) x_1\end{array}\right) \delta\alpha.
\end{equation}
where we suppressed the $\mathcal{O}(1/N^2)$ terms. The matrix on the left hand side is singular and the linear system has a solution if and only if the following constraint is satisfied:
\be
\label{delta alpha constraint}
4 \[ 3-8 \tilde g_2(3-16 \tilde g_3 x_1x_2) - 8 \tilde g_3 x_1 \big(1+4 \tilde g_3(1-2 x_1) \big) \]-x_1\,\delta\alpha = 0~.
\ee
This means  every fixed point of  the infinite $N$ conformal manifold \reef{branches}  survives finite $N$ corrections in a theory with the appropriate chosen $\alpha$. 

Let us demand that finite but large $N$ theory has $\alpha=1$. Corresponding fixed points are the intersections of  
 \reef{branches} with the following surface obtained by setting $\delta\alpha=0$ in (\ref{delta alpha constraint}), see Fig.~\ref{fig:infiniteN},
\be
 {3\over 8}= \tilde g_2(3-16 \tilde g_3 x_1x_2)  +   \tilde g_3 x_1 \big(1+4 \tilde g_3(1-2 x_1) \big)~. \label{surface}
\ee
One of the intersection points is $\tilde{g}_{1,2}=1/8, \tilde{g}_3=1/4$, which is the theory with full $O(N)$ symmetry.  Another fixed point 
has $g_3<0$, and exhibits persistent symmetry breaking, as discussed later in section \ref{sec: thermal physics}.
Similar conclusion holds for other values of $\alpha$. 

\noindent
Another class of tractable large-$N$ CFTs can be obtained by considering the limit $N\rightarrow \infty$ with fixed $m$.  In this case coupling constants will grow linearly with $N$,  $\tilde g_i = \tilde g_i^\mt{(1)} N + \tilde g_i^\mt{(0)}+\ldots$. Substituting this into \reef{beta-fn} and solving order by order in the large $N$ limit  yields a real solution of the form\footnote{As expected, in the limit $x_1=m/N\to 0$ the fixed points  $(\tilde g_1^+, \tilde g_2^+, \tilde g_3)$ of \reef{branches} converge to \reef{deriving alpha_c} for $m\gg 1$.}
\begin{equation}
\begin{aligned}
&\tilde g_1 = \frac{N}{8(m+8)}- 2\Big({g_3^\mt{(0)}(m+2)\over m+8}\Big)^2\,,\quad \tilde g_{2}=0 \,, \quad \tilde g_3= g_3^\mt{(0)}   \,, \quad\alpha = \frac{m+8}{m-4} ~, ~ m>4~,\\
&\tilde g_1 = \frac{N}{8(m+8)}- 72\Big({g_3^\mt{(0)}\over m+8}\Big)^2\,,\quad \tilde g_{2}={1\over 8} \,, \quad \tilde g_3= g_3^\mt{(0)}   \,, \quad\alpha = -\frac{m+8}{m-4} ~, ~ m<4\, ,
\end{aligned}
\label{deriving alpha_c}
\end{equation}
where the ${\cal O}(1/N)$ terms are suppressed. 
The solution is not unique, the coefficient $g_3^\mt{(0)}$
cannot be fixed without accounting for higher order corrections
in $1/N$. 
Moreover, $\alpha$ is constrained,
\begin{align}
\alpha =  \alpha_c \equiv \Big| \frac{m+8}{m-4}\Big|  ~.
 \label{crit-alpha}
\end{align}
Higher order in $1/N$ contributions remove the ambiguity in \reef{deriving alpha_c}.
Furthermore, higher order terms decrease the value of $\alpha$ such that $1\leq \alpha \leq \alpha_c$, \ie $\alpha_c$ is an upper critical value above which only complex fixed points exist. 
We verified explicitly there are fixed points with large and fixed $m$, $N\rightarrow \infty$, $\alpha\leq \alpha_c$ which have $\tilde g_3<0$ 
and which exhibit persistent symmetry breaking. 

\noindent
Fixed points for theories with small  $N\sim \mathcal{O}(1)$ are difficult to analyze analytically  but \reef{beta-fn} can be readily solved numerically. Ultimately, one finds a large family of stable fixed points with real-valued couplings and $\tilde g_3 < 0$. For $m=1$ there are such fixed points for any $N\geq 6$ and appropriate $\alpha$. Thus, persistent breaking of discrete global symmetry is possibly already in the $\Z_2\times O(5)$ model. In \cite{Chai:2021djc} it was reported that $N> 17$ is necessary for symmetry breaking  because the analysis there was restricted to $\alpha=1$ theories, while here we consider more general models with arbitrary positive $\alpha$. For $m=2$ fixed points with  $\tilde g_3 < 0$ (and persistent breaking) appear for all $N\geq 7$ and the appropriate $\alpha$. More generally for $m\leq 5$ we find  $\tilde g_3 < 0$ fixed points for all  $N\geq m+5$ and for $m\ge 5$ for all  $N\geq 2m$, using the  nomenclature $m\leq N-m$.

We note that negative $\tilde g_3$  is necessary but not sufficient for persistent symmetry breaking. So far our emphasis was on the value of $\tilde g_3$. We will consider finite temperature effects in more detail in section~\ref{sec: thermal physics}.

\subsection{Family of interacting CFTs at finite $N$}
\label{sec:confmanifold}

In section~\ref{sec:model}, while working in the infinite  $N$ limit, 
we  encountered a conformal manifold
of interacting fixed points. In this section, we are going
to discuss the origin of this conformal manifold, as well as
its fate at finite $N$.

We are interested in analyzing fixed points of the model (\ref{Sdeform}).
At the Gaussian fixed point, $g_i = 0$, $i=1,2,3$,
we have a continuous family of CFTs,
parametrized by scaling dimensions $\Delta_{\phi_j}$, $j=1,2$.
In the previous section, we expressed these scaling dimensions as $\Delta_{\phi_j}=(d-\epsilon_j)/4$,
in terms of the ratio $\alpha = \epsilon _1/\epsilon_2$, and $\epsilon_2$. This 
is convenient to do because in interacting theory, while working
at the linear order in perturbation theory (in $\epsilon_{1,2}$) one can simply factor $\epsilon_2$
out of all expressions. 

Let us  fix $\epsilon_i$ in the UV. Since \eqref{S0} are non-local, they are not getting renormalized, and therefore values of $\Delta_{\phi_1}, \Delta_{\phi_2}$ remain the same along the RG flow. Now let us assume this theory admits a non-trivial interacting fixed point  $(g_1^\star,g_2^\star,g_3^\star)$.
There are different scenarios concerning whether it belongs to a continuous family:
\begin{itemize}
\item
Fixed point $(g_1^\star,g_2^\star,g_3^\star)$ is a solution of equations $\beta_i = 0$, $i=1,2,3$, and it exists for some open set of  $\Delta_{\phi_j}$, $j=1,2$.
Then corresponding  interacting fixed points form a  continuous family,  parametrized by $\Delta_{\phi_j}$, $j=1,2$,
just like it was for  the Gaussian fixed points. This is the general case scenario, which assumes the fixed point equations $\beta_i(g_j,\epsilon_k)=0$ are non-degenerate. 
\item
Fixed point $(g_1^\star,g_2^\star,g_3^\star)$  is a solution of $\beta_i = 0$,  that exists only
 for the  isolated point(s) on the $(\Delta_{\phi_1},\Delta_{\phi_2})$
plane. There is no continuous family of interacting fixed points in this case.
\item 
The system of equations $\beta_i = 0$,  is degenerate w.r.t.~$g_{1,2,3}$, and for consistency $\Delta_{\phi_1},\Delta_{\phi_2}$ must be related to each other, forming a curve  on the $(\Delta_{\phi_1},\Delta_{\phi_2})$
plane. There is  a continuous family of interacting fixed points in this case,  it is two-dimensional, and intersects
the $(\Delta_{\phi_1},\Delta_{\phi_2})$ plane along a one-dimensional curve.
\end{itemize}
In each scenario above, besides $\beta_i(g^*)=0$, we also assumed $g_i^*$ satisfy additional constraints, e.g.~$g_{1,2}^*>0$, to ensure stability of the model. 

In section~\ref{sec:model}, we saw that  in the infinite $N$ limit we have the third scenario, while  finite $N$ models follow first scenario, i.e.~also admit a continuous family of fixed points. 
Specifically, in the infinite $N$ limit, for fixed $m/N={\cal O}(1)$, we 
saw that the system of quadratic  equations (\ref{RGinfN}) is degenerate. Its admits a one-parametric family of solutions (\ref{branches}). The corresponding exactly marginal operator ${\cal O}_+'$,
that we discuss in more detail latter in 
section~\ref{sec:anomalous dimensions},
manifests existence of a one-dimensional conformal manifold. Together with the
parameter $\epsilon_1=\epsilon_2$, and associated non-local deformation of \eqref{S0}, we obtain a two-dimensional continuous family of CFTs.
It is algebraically straightforward to find $1/N$ corrections to the
leading order solution  (\ref{branches}). For instance, at the next-to-leading order
one needs to solve the linearized system of equations
(\ref{subleading 1/N g's}).
The one-dimensional conformal manifold in the $\epsilon_1=\epsilon_2$ theory, that we observed in the infinite $N$
limit, is lifted by the $1/N$ corrections: there is no marginal  operator ${\cal O}'$
(for any possible mixing of the double-trace operators ${\cal O}_i$)
in this case. However, imposing the constraint (\ref{delta alpha constraint})
one is able to see that by changing the parameters $\epsilon_{1,2}$ in the UV the two-parametric family of interacting CFTs survives. A similar
argument will remain true at any order in $1/N$ expansion.

In fact, it is easy to see numerically that the two-parametric family of interacting CFTs exists at finite $N$.  For instance, in Fig.~\ref{fig:m1N6} for the theory with  $m=1$, $N=6$ we plot 
critical $\tilde g_3^\star$ for various admissible values of $\alpha$ for which interacting
fixed point exists. Analogous numerical analysis
can be used to establish existence of the family of interacting CFTs for general $m,N$.\footnote{
Interestingly enough, there is the aforementioned family for $m=1$ and any $N\geq 2$, except for $N=4$, when  the continuous family of solutions 
disappears. Solutions still seem to exist for a certain discrete set of $\alpha$,  realizing
second scenario above (although $\epsilon_1$ is still continuous). For example, for $\alpha=1$, $N=4$, $m\in \{1,2,3\}$,
we obtain $\tilde g_1=\tilde g_2=\frac{1}{24}$, $\tilde g_3 = \frac{1}{12}$. This solution,
however, has positive $g_3$ and hence does not exhibit the PSB.}

\subsection{Anomalous dimensions}
\label{sec:anomalous dimensions}
Before we proceed with the finite temperature analysis in section \ref{sec: thermal physics} we would like to provide some checks that the fixed points with negative $g_3$, which we found in section \ref{sec:model}, correspond to UV complete unitary theories. For that purpose in this section we calculate  anomalous dimensions of the composite operators ${\cal O}_i$\footnote{In this paper, we denote quartic and quadratic operators in the action as $\mathcal{O}_i$ and $\tilde{\mathcal{O}_i}$ respectively, while  the analogous operators at the interacting fixed points are denoted with primes.}. We find the anomalous dimensions to be real, which is a necessary condition for unitarity. 

We begin with repeating the derivation of the anomalous dimensions at leading order of the conformal perturbation theory  \cite{Behan:2017emf,Komargodski:2016auf, Cardy:1996xt, Cappelli:1991ke}, keeping in mind that our theory is non-local.
At linear order the conformal perturbation theory gives
\be
 \langle \mathcal{O}_i(x_1) \mathcal{O}_j(x_2)\rangle = \delta_{ij} \, {N_i\over |x|^{2\Delta_i}} - 
  \sum_{k=1}^3 {g_k \mu^{\epsilon_k} \over N}  \int d^d x_3 
  \langle \mathcal{O}_i(x_1) \mathcal{O}_j(x_2) \mathcal{O}_k(x_3)\rangle + \mathcal{O}(g_k^2) ~,
\ee
where the three point function is calculated at the Gaussian fixed point. After  substituting the leading order OPE expansion, 
\be
 \mathcal{O}_i(x_1) \mathcal{O}_k(x_3) = \sum_{j=1}^3 {C_{ik}^j \over |x_{13}|^{\Delta-2\Delta_j}} \mathcal{O}_j(x_1) + \ldots ~,
\ee
we get
\be
 \langle \mathcal{O}_i(x_1) \mathcal{O}_j(x_2)\rangle = \delta_{ij} \, {N_i\over |x_{12}|^{2\Delta_i}} - 
  \sum_{k=1}^3 {g_k \mu^{\epsilon_k} \over N} \Big( {C_{ik}^j N_j\over |x_{12}|^{2\Delta_j}} 
  \int  {d^d x_3 \over |x_{3}|^{\Delta-2\Delta_j}} +(i \leftrightarrow j) \Big)+ \mathcal{O}(g_k^2)~.
\ee
Integrating out within a shell $\mu^{-1} < |x_3| < \mu^{-1}_\mt{IR}$ between the subtraction scale $\mu$ and the IR cutoff $\mu_{IR}$ results in the following change
\be
 \delta \langle \mathcal{O}_i(x_1) \mathcal{O}_j(x_2)\rangle = - 
   {2\pi^{d\over 2} \over \Gamma\({d\over 2}\)} 
   \sum_{k=1}^3 {g_k \mu^{\epsilon_k} \over N}
    \Big({C_{ik}^j N_j\over |x_{12}|^{2\Delta_j}} {\mu_\mt{IR}^{-\epsilon_{ikj}} - \mu^{-\epsilon_{ikj}}\over \epsilon_{ikj} }+ (i \leftrightarrow j) \Big)+ \mathcal{O}(g_k^2) ~,
\ee
where $\epsilon_{ikj} = \epsilon_i+\epsilon_k - \epsilon_j$. 
Consider the case of equal epsilons, $\epsilon_i=\epsilon$. Then
\be
 \mu {\del \over \del \mu}\langle \mathcal{O}_i(x_1) \mathcal{O}_j(x_2)\rangle = 
 -{2 \epsilon\over N \, |x_{12}|^{2\, (d-\epsilon)} }  \sum_{k=1}^3  \tilde g_k \big( C_{ik}^j N_j +  C_{jk}^i N_i  \big) 
 + \mathcal{O}(\epsilon^2) ~.
 \label{CZ}
\ee
By $\mathcal{O}'_i$ we will denote primary operators in the weakly interacting theory at the fixed point. 
The Callan-Symanzyk equation for  $\mathcal{O}'_i$ with the anomalous dimension $\gamma_i$ is given by
\be
\mu {\del \over \del \mu}\langle \mathcal{O}'_i(x_1) \mathcal{O}'_i(x_2)\rangle = -2 \gamma_i \, \langle \mathcal{O}'_i(x_1) \mathcal{O}'_i(x_2)\rangle ~,
\label{CZgen}
\ee
where to zeroth order in $\epsilon$
\be
\label{CZ1}
 \mathcal{O}'_i =  \sum_{k=1}^3 V_i^k \, \mathcal{O}_k  + \mathcal{O}(\epsilon) ~, \quad
 \langle \mathcal{O}'_i(x_1) \mathcal{O}'_j(x_2)\rangle =  {\sum_{k=1}^3V_i^k V_j^k N_k \over |x_{12}|^{2\, (d-\epsilon)} } 
 ={\delta_{ij} \over |x_{12}|^{2\, (d-\epsilon)} } 
 ~.
\ee
The transition matrix $V_i^k$ is determined by requiring compatibility of \reef{CZgen} with  \reef{CZ1}, \ie using \reef{CZ1} and definition of $\mathcal{O}'_i$, we obtain
\bea
&& \mu {\del \over \del \mu}\langle \mathcal{O}'_m(x_1) \mathcal{O}'_n(x_2)\rangle = 
 -{1\over  |x_{12}|^{2\, (d-\epsilon)} }  \sum_{i,j}  \(N_j V^j_n \gamma_{ji} V^i_m + N_i V^i_m \gamma_{ij} V^j_n  \) 
 + \mathcal{O}(\epsilon^2) \, ,
 \label{CZ2}
 \\
&&  \gamma_{ji} = {2 \epsilon\over N} \sum_k  \tilde g_k  C_{ik}^j = {\del\beta_j\over \del g_i} + \epsilon \delta_{ij} ~.
\nonumber
\eea
Let us choose $V^i_m$ to be the eigenvectors of $\gamma_{ji}$ with eigenvalues $\gamma_m$, or equivalently, $\sum_{i=1}^3\gamma_{ji}V^i_m = \gamma_m V^j_m$, then\footnote{Two terms within parenthesis in \reef{CZ2} are equal, because it follows from \reef{OPErelation} that $N_j \gamma_{ji} =  {2 \epsilon\over N} \sum_k  \tilde g_k  C_{ij}^k N_k$ is a symmetric matrix. Hence, for our choice of $V^i_m$ we get $\gamma_m \sum_j N_j V^j_n V^j_m = \gamma_n \sum_i N_i V^i_n V^i_m$, and therefore $\sum_j N_j V^j_n V^j_m = \delta_{mn}$, because degeneracy is lifted, \ie $\gamma_m\neq \gamma_n$ for $m\neq n$.}
\be
\mu {\del \over \del \mu}\langle \mathcal{O}'_m(x_1) \mathcal{O}'_n(x_2)\rangle =
 -{2\gamma_m \delta_{mn}\over  |x_{12}|^{2\, (d-\epsilon)} }  
 + \mathcal{O}(\epsilon^2) \, .
\ee
Comparing to \reef{CZgen}, we conclude that the anomalous dimensions are given by the eigenvalues of $\gamma_{ij}$.

The derivation of anomalous dimensions for non-equal $\epsilon_i$'s is similar. This is a nearly degenerate case, and therefore it is convenient to introduce an intermediate $\epsilon$ defined by $\epsilon_i=\epsilon + \delta\epsilon_i$ with $\delta\epsilon_i\sim \epsilon$. Next we rescale the operators $\mathcal{O}_i \to \mu^{\delta\epsilon_i}  \mathcal{O}_i$. In particular, \reef{CZ} for {\it rescaled} fields takes the form
\be
 \mu {\del \over \del \mu}\langle \mathcal{O}_i(x_1) \mathcal{O}_j(x_2)\rangle = 
 { - 2 \over |x_{12}|^{2\, (d-\epsilon)} }\Big( - \delta\epsilon_j \, N_j \,\delta_{ij}  + {1\over N} \sum_{k=1}^3  \epsilon_k \tilde g_k \big( C_{ik}^j N_j +  C_{jk}^i N_i  \big) \Big)
 + \mathcal{O}(\epsilon^2) ~.
 \label{rescaledCZ}
\ee
Repeating the same steps as before, we conclude that the anomalous dimensions of $\mathcal{O}'_m$ in the nearly degenerate case are given by the eigenvalues, $\gamma_m$, of the matrix
\be
 \gamma_{ij} = - \delta\epsilon_i  \,\delta_{ij}  + {2 \over N} \sum_k  \epsilon_k \tilde g_k  C_{jk}^i = {\del\beta_i\over \del g_j} + \epsilon \delta_{ij} ~.
\ee

\noindent
To recapitulate, we find that non-locality of the model does not affect the expression for the anomalous  dimensions. At leading order of the conformal perturbation theory \cite{Behan:2017emf,Komargodski:2016auf,Cardy:1996xt, Cappelli:1991ke} they are given by the eigenvalues of the derivatives matrix  ${\del\beta_i\over \del g_j}$ evaluated at the fixed point.  For the scaling dimensions of the  operators $\mathcal{O}'_i$ we find
\be
 \Delta_m'=d - \epsilon + \gamma_m = d + \omega_m ~,
\ee 
where $\omega_m$ are the eigenvalues of the derivatives matrix  ${\del\beta_i\over \del g_j}$.

\noindent
The anomalous dimensions simplify in the case when  all epsilons are equal.  Then one of the three eigenvalues of $\gamma_{ij}$ can be readily derived using $\beta_i=0$ and \reef{RGflow},
\be
 \sum_{j=1}^3 \gamma_{ij} \tilde g_j = 2\epsilon \tilde g_i ~.
\ee
This eigenvalue corresponds to an irrelevant operator $\mathcal{O}'=\sum_{i=1}^3 \tilde g_i \mathcal{O}_i$ with scaling dimension 
\be
 \Delta'=d+\epsilon ~. 
\ee
The scaling dimensions of two additional operators, $\mathcal{O}'_\pm$ ,are given by
\bea
&& \Delta'_\pm=d-\epsilon + \gamma_\pm~,
\\
 &&\gamma_\pm={4\epsilon\over N}\( \kappa_2 \pm \sqrt{\kappa_2^2 - \kappa_1} \) \,, ~ 
  \kappa_1 =48(N+16) \tilde g_1 \tilde g_2\, , ~ 
  \kappa_2= \tilde g_1(m+14) + \tilde g_2(N-m+14)~.
  \nonumber
\eea
Note that they are real, because the radicand in the above expression is non negative,
\be
 \kappa_2^2 - \kappa_1=g_1^2(m+14)^2+g_2^2(N-m+14)^2-2g_1 g_2\big[ (m+14)(N-m+14) - 2 (m+2)(N-m+2)\big]~,
\ee
and therefore,\footnote{Stability of the fixed point requires $g_1$ and $g_2$ to be non-negative.}
\be
 \kappa_2^2 - \kappa_1\geq \big[g_1(m+14)- g_2 (N-m+14) \big]^2 \geq 0 ~.
\ee 
For finite $N$ and $m$ the operators with scaling dimensions $\Delta'_+$ and $\Delta'_-$ are weakly irrelevant and relevant respectively. Thus the critical surface (subspace of irrelevant deformations at the interacting fixed point) has codimension 1 in the space of nearly marginal couplings $(g_1, g_2, g_3)$.  

\noindent
Using the last equation in \eqref{beta-fn}, we deduce that in the large rank limit with $m/N$ fixed $\Delta_+\to d$, $\Delta_-\to d-\epsilon$. The $\mathcal{O}'_+$ operator corresponds to an exactly marginal deformation associated with a line of fixed points (\ref{branches}) that emerges in the infinite $N$ limit. Each point on the conformal manifold has one weakly relevant and one weakly irrelevant deformation with scaling dimensions $d\mp\epsilon$ respectively.

Next consider the following single trace operators
\be
 \widetilde{\mathcal{O}}_1= \phi_1^2 ~, \widetilde{\mathcal{O}}_2= \phi_2^2 ~.
\ee
We remind the reader that by ${\cal O}_i$ we denote quartic operators, invariant under $O(m)\times O(N-m)$, while $\tilde{\cal O}_i$ above are quadratic in fields.
Their scaling dimensions at the Gaussian fixed point are given by $\widetilde\Delta_1=2\Delta_{\phi_1}$, $\widetilde\Delta_2=2\Delta_{\phi_2}$. The two-  and three-point functions satisfy
\bea
 && \langle \widetilde{\mathcal{O}}_j(x) \widetilde{\mathcal{O}}_k(0)\rangle = \delta_{jk} \, {\widetilde N_k\over |x|^{2\widetilde\Delta_k}} ~,
 \non
&& \widetilde N_1 = 2 m~, \quad \widetilde N_2 = 2 (N-m) ~,
 \label{2p}
\eea
and
\bea
\nonumber
\langle \mathcal{O}_i(x_1) \widetilde{\mathcal{O}}_j(x_2) \widetilde{\mathcal{O}}_k(x_3)\rangle&=&
{\widetilde C_{ij}^k \widetilde{N}_k \over
|x_{12}|^{\widetilde\Delta-2\widetilde\Delta_k} 
|x_{23}|^{\widetilde\Delta-2\Delta_i} 
|x_{13}|^{\widetilde\Delta-2\widetilde\Delta_j}},\\
\widetilde\Delta&=&\Delta_i+\widetilde\Delta_j+\widetilde\Delta_k,
\eea
where the non-zero OPE coefficients $\widetilde C^{k}_{ij}$ can be expressed in terms of $C^k_{ij}$
\bea
 && \widetilde C_{11}^1= C^3_{13}~, ~ \widetilde C_{31}^2=C^2_{33}~, ~ \widetilde C^2_{22}=C^3_{32}~,~ \widetilde C^1_{32}= C^1_{33} ~.
\eea
As before they are related by
\be
\widetilde C_{ij}^k = \widetilde C_{ik}^j \widetilde N_j/ \widetilde N_k ~ .
\label{OPErelation2}
\ee

\noindent
To calculate the leading order correction to the scaling dimensions of  $\widetilde{\mathcal{O}}_i$ at the weakly interacting fixed point, we resort to the linear order conformal perturbation theory 
\be
 \langle \widetilde{\mathcal{O}}_j(x_2) \widetilde{\mathcal{O}}_k(x_3)\rangle = \delta_{jk} \, 
 {N_k\over |x_{23}|^{2\widetilde\Delta_k}} - 
  \sum_{i=1}^3 {g_i \mu^{\epsilon_i} \over N}  \int_{\mu^{-1}} d^d x_1 
  \langle  \mathcal{O}_i(x_1) \widetilde{\mathcal{O}}_j(x_2) \widetilde{\mathcal{O}}_k(x_3) \rangle + \mathcal{O}(g_k^2) ~,
\ee
where the three point function is calculated at the Gaussian fixed point, and $\mu$ is a floating cutoff scale. In particular, using the leading order OPE expansion, 
\be
 {\mathcal{O}}_i(x_1) \widetilde{\mathcal{O}}_j(x_2) = \sum_{k=1}^2 
 {\widetilde C_{ij}^k \over |x_{12}|^{\widetilde\Delta-2\widetilde\Delta_k}} \widetilde{\mathcal{O}}_k(x_2) + \ldots ~,
 \label{OPE}
\ee
one can calculate a small change in the two point function under variations in the floating cutoff scale $\mu$,
\be
 \delta \langle \widetilde{\mathcal{O}}_j(x_2) \widetilde{\mathcal{O}}_k(x_3)\rangle = - 
  \sum_{i=1}^3 {g_i \mu^{\epsilon_i} \over N} \Big( {\widetilde C_{ij}^k  \widetilde N_k\over |x_{23}|^{2\widetilde\Delta_k}} 
  \int_{\mu^{-1}}^{\mu^{-1}_\mt{IR}}  {d^d x_1 \over |x_{1}|^{\widetilde\Delta-2\widetilde\Delta_k}} +(k \leftrightarrow j) \Big)+ \mathcal{O}(g_k^2)~.
\ee

\noindent
In the case of equal epsilons, $\epsilon_i=\epsilon$, we obtain
\be
 \mu {\del \over \del \mu}  \langle \widetilde{\mathcal{O}}_j(x_2) \widetilde{\mathcal{O}}_k(x_3)\rangle = 
 -{2 \epsilon\over N \, |x_{23}|^{d-\epsilon} }  \sum_{i=1}^3  \tilde g_i \big( \widetilde C_{ij}^k \widetilde N_k + \widetilde C_{ik}^j \widetilde N_j  \big) 
 + \mathcal{O}(\epsilon^2) ~.
\ee

\noindent
At this point we introduce primary operators in the weakly interacting CFT, \ie $\widetilde{\mathcal{O}}'_i=\sum_{k=1}^2 \widetilde V_i^k \, \widetilde{\mathcal{O}}_k+ \mathcal{O}(\epsilon) $  with the anomalous dimension $\widetilde \gamma_i$. 
The Callan-Symanzyk equation for  $\widetilde{\mathcal{O}}'_i$  is given by
\be
\mu {\del \over \del \mu}\langle \widetilde{\mathcal{O}}'_i(x_1) \widetilde{\mathcal{O}}'_i(x_2)\rangle 
= -2 \widetilde\gamma_i \, \langle \widetilde{\mathcal{O}}'_i(x_1) \widetilde{\mathcal{O}'}_i(x_2)\rangle ~,
\label{CZphi2}
\ee
where
\be
\langle \widetilde{\mathcal{O}}'_i(x_1) \widetilde{\mathcal{O}}'_j(x_2)\rangle =  
{\sum_{k=1}^2 \widetilde V_i^k \widetilde V_j^k \widetilde N_k \over |x_{12}|^{d-\epsilon} } 
 ={\delta_{ij} \over |x_{12}|^{d-\epsilon} } 
 ~.
 \label{CZgen2}
\ee
The transition matrix $\widetilde V_i^k$ is determined by requiring compatibility of \reef{CZphi2} with \reef{CZgen2}, 
\bea
&& \mu {\del \over \del \mu}\langle \widetilde{\mathcal{O}}'_m(x_1) \widetilde{\mathcal{O}}'_n(x_2)\rangle = 
 -{1\over  |x_{12}|^{d-\epsilon} }  \sum_{k,j=1}^2  \(\widetilde N_k \widetilde V^k_n \widetilde \gamma_{kj} \widetilde V^j_m + \widetilde N_j \widetilde V^j_m \widetilde \gamma_{jk} \widetilde V^k_n  \) 
 + \mathcal{O}(\epsilon^2) \, ,
 \label{2CZphi2}
 \\
&& \widetilde \gamma_{kj} = {2 \epsilon\over N} \sum_{i=1}^3  \tilde g_i  \widetilde C_{ij}^k ~.
\nonumber
\eea
Let us choose $\widetilde V^j_m$ to be the eigenvectors of $\widetilde \gamma_{kj}$ with eigenvalues $\widetilde \gamma_m$, or equivalently, $\sum_{j=1}^2\widetilde \gamma_{kj}\widetilde V^j_m = \widetilde \gamma_m \widetilde V^k_m$, then\footnote{Two terms within parenthesis in \reef{2CZphi2} are equal, because it follows from \reef{OPErelation2} that $\widetilde N_j \widetilde\gamma_{ji}$ is a symmetric matrix. Hence, for our choice of $\widetilde V^i_m$ we get $\widetilde \gamma_m \sum_k \widetilde N_k \widetilde V^k_n \widetilde V^k_m = \widetilde \gamma_n \sum_j \widetilde N_j \widetilde V^j_n \widetilde V^j_m$, and therefore $\sum_k \widetilde N_k \widetilde V^k_n \widetilde V^k_m = \delta_{mn}$, because degeneracy is lifted, \ie $\widetilde \gamma_m\neq \widetilde \gamma_n$ for $m\neq n$.}
\be
\mu {\del \over \del \mu}\langle \widetilde{\mathcal{O}'}_m(x_1) \widetilde{\mathcal{O}}'_n(x_2)\rangle =
 -{2\widetilde\gamma_m \delta_{mn}\over  |x_{12}|^{d-\epsilon} }  
 + \mathcal{O}(\epsilon^2) \, .
\ee
Comparing to \reef{CZgen}, we conclude that the anomalous dimensions are given by the eigenvalues of $ \widetilde \gamma_{kj}$,
\bea
 \widetilde\gamma_\pm&=&{\epsilon\over N} \(C^3_{13} \tilde g_1 + C^3_{32} \tilde g_2 
 \pm \sqrt{\big(C^3_{13} \tilde g_1 - C^3_{32}\tilde g_2\big)^2 + 4 C^1_{33} C^2_{33} \tilde g_3^2}  \) ~,
 \nonumber \\
 \widetilde V_\pm &=& \(
 { C^3_{13} \tilde g_1-C^3_{32}\tilde g_2
 \pm \sqrt{\big(C^3_{13} \tilde g_1 - C^3_{32} \tilde g_2\big)^2 + 4 C^1_{33} C^2_{33} \tilde g_3^2} \over 
 2 C^2_{33} \tilde g_3} ~~, ~~ 1 ~ \) ~. \label{reals}
\eea
Note that the anomalous dimensions, $\widetilde\gamma_\pm$, are manifestly real for all $m$ and $N$.

\noindent
For non-equal $\epsilon_i$'s, the scaling dimensions of two operators $\widetilde{\mathcal{O}}'_i$ can be similarly derived
\be
 \widetilde \Delta_m'={d\over 2} + \widetilde \omega_m ~,
\ee
where $\widetilde \omega_m$ are the eigenvalues of the $2\times 2$ matrix
\be
\label{adim}
 \widetilde \gamma_{kj} = - {\epsilon_k \over 2} \delta_{kj} 
 + {2 \over N} \sum_{i=1}^3 \epsilon_i \tilde g_i  \widetilde C_{ij}^k ~.
\ee
In the vicinity of $\alpha=1$, i.e.~for $\epsilon_1\approx \epsilon_2$ the eigenvalues of \eqref{adim} are perturbatively close to \eqref{reals}, which are real. Therefore eigenvalues of \eqref{adim}  will also be real, at least until two of them collide. 

\noindent To conclude, we have calculated anomalous dimensions of all quadratic and quartic operators at the interacting fixed points, at leading order in the conformal perturbation theory. Assuming the fixed point is stable, all scaling dimensions are real, 
as it is necessary for unitarity of the IR theory.

\subsection{Tests of conformal invariance}
\label{sec:tests of conformal invariance}

The fixed point QFT is certainly scale invariant, but it is not necessarily a CFT. In this section we perform a number of tests to provide  evidence that the scale symmetry in our models is enhanced to the full conformal group.  For simplicity we consider the case of equal epsilons only, $\epsilon_i=\epsilon$. The case of non equal epsilons is similar. 

\noindent
We start by calculating the two point function of primaries at the fixed point, $\langle \widetilde{\mathcal{O}'}_i\mathcal{O}'_j\rangle$. These correlators could be  non-zero if the model is scale invariant but non-conformal. However, it must vanish if the theory exhibits full conformal symmetry. 
\begin{figure}
\begin{center}
\includegraphics[width=0.3\textwidth]{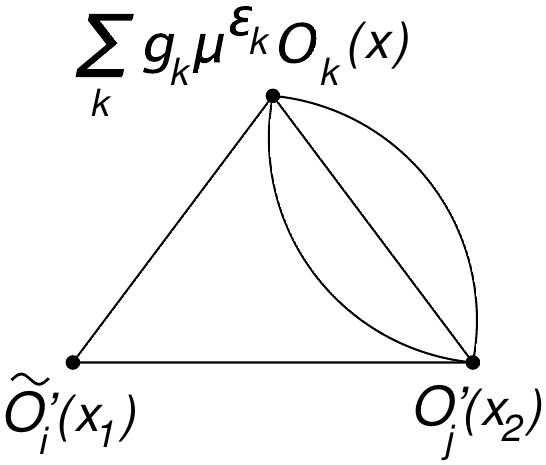}
\caption{Linear order contribution to the correlator $\langle \widetilde{\mathcal{O}'}_i(x_1)\mathcal{O}'_j(x_2)\rangle$. Solid lines represent propagators of the scalar fields. Integration is done over the insertion point $x$.}
\label{fig:2pTest}
\end{center}
\end{figure}

\noindent
To linear order in $\epsilon$ to evaluate $\langle \widetilde{\mathcal{O}'}_i\mathcal{O}'_j\rangle$ we should calculate the  diagram shown in Fig. \ref{fig:2pTest}. Up to an overall factor, we have
\be
 \text{Fig.}\ref{fig:2pTest}\propto {\epsilon \, \mu^\epsilon\over |x_{12}|^{d-\epsilon\over 2}}\int d^d x {1\over |x-x_1|^{d-\epsilon\over 2} |x-x_2|^{3(d-\epsilon)\over 2}} 
 =\pi^{d\over 2}  \, {\epsilon \, (|x_{12}|\mu)^\epsilon\over |x_{12}|^{3(d-\epsilon)\over 2}}
 {\Gamma\big({d-2\epsilon\over 2}\big)\Gamma\big({d+\epsilon\over 4}\big)\Gamma\big({3\epsilon-d\over 4}\big)
 \over \Gamma\big({d-\epsilon\over 4}\big)\Gamma\big({3(d-\epsilon)\over 4}\big)\Gamma(\epsilon)} \label{fig2}
 ~,
\ee
where the proportionality constant is some function of $\tilde g_k$, $N$ and $m$, and we used the following identity
\bea
\int d^d x \frac{1}{ |x-x_1|^{\alpha} |x-x_2|^\beta}
= \pi^{d/2} {\Gamma\big({\al+\bt-d\over 2}\big) \over
 \Gamma(\al/2)\Gamma(\bt/2)} {\Gamma\({d-\al\over 2}\)\Gamma\big({d-\beta \over 2}\big) \over \Gamma(d-\al/2-\bt/2)} |x_{12}|^{d-\al-\bt}~.
 \label{identity}
\eea
Gamma function $\Gamma(\epsilon)$ in the denominator of \eqref{fig2} introduces one extra power of $\epsilon$ and therefore  to linear order in $\epsilon$ that expression vanishes, yielding
\be
\langle \widetilde{\mathcal{O}'}_i(x_1)\mathcal{O}'_j(x_2)\rangle =0+\mathcal{O}(\epsilon^2) ~.
\ee
As can be seen from \reef{identity}, this result also holds if $\mathcal{O}'_j(x_2)$ is replaced with a scalar operator which is quartic in fields and has any number of derivatives. Likewise the linear order correction to the correlation function of $\widetilde{\mathcal{O}'}_i$ with any operator which is more than quartic in fields  also  vanishes identically. 

\begin{figure}
\begin{center}
\includegraphics[width=0.4\textwidth]{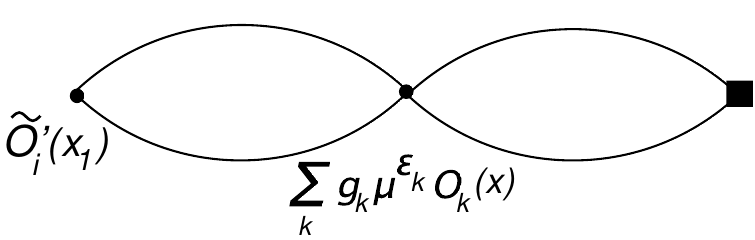}
\caption{Linear order contribution to the correlation function of $\widetilde{\mathcal{O}'}_i$ with a primary built off two scalar fields and any number of derivatives (solid square). Solid lines represent propagators of the scalar fields. Integration is done over the insertion point $x$.}
\label{fig:2pTest2}
\end{center}
\end{figure}

\noindent
Consider now the correlation function of $\widetilde{\mathcal{O}'}_i$ with a primary operator, which is  quadratic in fields and which includes any number of derivatives. This correlator vanishes to zeroth order in the coupling constant, because the Gaussian theory is conformal. The linear order correction is shown in Fig. \ref{fig:2pTest2}, it factorizes into a product of two sub-diagrams. One of them vanishes, because it is proportional to the correlation function of two distinct primaries in the Gaussian model. 

\noindent
The upshot of this calculation is that up to linear order in $\epsilon$ the two point function of $\widetilde{\mathcal{O}'}_i$ with various operators is compatible with the  conformal symmetry -- all two-point correlators of operators with non-equal dimensions vanish. This result can be used to show that the three point function, 
\be
 \langle \widetilde{\mathcal{O}'}_i(x_1)\widetilde{\mathcal{O}'}_j(x_2)\widetilde{\mathcal{O}'}_k(x_3)\rangle = 
 \sum_{\ell_1,\ell_2,\ell_3=1}^2\widetilde V_i^{\ell_1}\widetilde V_j^{\ell_2} \widetilde V_k^{\ell_3}
 \langle\widetilde{ \mathcal{O}}_{\ell_1}(x_1) \widetilde{\mathcal{O}}_{\ell_2}(x_2) \widetilde{\mathcal{O}}_{\ell_3}(x_3)\rangle~.
 \label{3p-function}
\ee
is conformal up to linear order in $\epsilon$.

\noindent
Indeed, integrating out distances within the shell $\mu^{-1} < |x| < \mu^{-1}_\mt{IR}$ and using \reef{OPE} results in the following change
\bea
&&  \delta \langle\widetilde{ \mathcal{O}}_{\ell_1}(x_1) \widetilde{\mathcal{O}}_{\ell_2}(x_2) \widetilde{\mathcal{O}}_{\ell_3}(x_3)\rangle = 
     \\
&&    -  \sum_{k=1}^3 {g_k \mu^{\epsilon} \over N} \Big( \sum_{m=1}^2 \tilde C_{k \ell_3}^m  
\langle\widetilde{ \mathcal{O}}_{\ell_1}(x_1) \widetilde{\mathcal{O}}_{\ell_2}(x_2) \widetilde{\mathcal{O}}_m(x_3)\rangle_0
  \int_{\mu^{-1}}^{\mu^{-1}_\mt{IR}}  {d^d x \over |x|^{d-\epsilon}} +(1\leftrightarrow 3, \, 2\leftrightarrow 3) \Big)+ \mathcal{O}(g_k^2)~.
  \nonumber
\eea
Or equivalently, 
\bea
  \mu {\del \over \del \mu}  \langle\widetilde{ \mathcal{O}}_{\ell_1}(x_1) \widetilde{\mathcal{O}}_{\ell_2}(x_2) \widetilde{\mathcal{O}}_{\ell_3}(x_3)\rangle &=& 
  - \sum_{m=1}^2 \tilde\gamma_\mt{m$\ell_3$}  
\langle\widetilde{ \mathcal{O}}_{\ell_1}(x_1) \widetilde{\mathcal{O}}_{\ell_2}(x_2) \widetilde{\mathcal{O}}_m(x_3)\rangle_0
   \nonumber
   \\
   &+&(1\leftrightarrow 3, \, 2\leftrightarrow 3) + \mathcal{O}(g_k^2) ~,
\eea
where $\tilde\gamma_\mt{m$\ell_3$} $ is defined in (\ref{2CZphi2}). Plugging it back into \reef{3p-function} and using the fact that $V^\ell_i$ are the eigenvectors of $\tilde\gamma_\mt{m$\ell$} $ with eigenvalues $\widetilde \gamma_i$, we conclude that the three point function \reef{3p-function} takes the following general form
\be
 \langle \widetilde{\mathcal{O}'}_i(x_1)\widetilde{\mathcal{O}'}_j(x_2)\widetilde{\mathcal{O}'}_k(x_3)\rangle \sim 
 \sum { \mu^{-\tilde\gamma_i-\tilde\gamma_j-\tilde\gamma_k}\over 
 |x_{12}|^{{d-\epsilon\over 2} + \alpha_{ijk}} |x_{13}|^{{d-\epsilon\over 2} + \alpha_{ikj}} |x_{23}|^{{d-\epsilon\over 2} + \alpha_{jki}}}~,
\ee
where the sum includes all possible $\alpha$'s which satisfy $\alpha_{ijk}=\alpha_{jik}\sim \epsilon$ and $\alpha_{ijk}+\alpha_{ikj}+\alpha_{jki}=\tilde\gamma_i + \tilde\gamma_j + \tilde\gamma_k$. In particular, for $i=j=k$, we get only one possible term with $\alpha_{iii}=\gamma_i$, and the associated three point function at linear order in $\epsilon$ is necessarily conformal. We should only consider the case when one of the three indices $i,j,k$ is different from the other two, \eg $i=j=+$ and $k=-$. In the limit $x_1\to x_2$ the leading order singularity takes the form
\be
 \langle \widetilde{\mathcal{O}'}_+(x_1)\widetilde{\mathcal{O}'}_+(x_2)\widetilde{\mathcal{O}'}_-(x_3)\rangle 
 \underset{x_1\sim x_2}{\longrightarrow} 
  { \mu^{-\tilde\gamma_i-\tilde\gamma_j-\tilde\gamma_k}\over 
 |x_{12}|^{{d-\epsilon\over 2} + \alpha_{++-}} |x_{13}|^{d-\epsilon + 2\alpha_{+-+}} }~.
\ee
Since $x_1\sim x_2$, one can substitute an appropriate OPE for the $\widetilde{\mathcal{O}'}_+(x_1)\widetilde{\mathcal{O}'}_+(x_2)$ on the left hand side. However, we have shown that the two-point function of $\widetilde{\mathcal{O}'}_-$ with various operators respects conformal symmetry to linear order in $\epsilon$. Thus, only the term proportional to $\widetilde{\mathcal{O}'}_-$ in the OPE contributes in this limit. As a result, the left hand side scales as $1/|x_{13}|^{d-\epsilon+2\gamma_-}$. In particular, $\alpha_{+-+}=\gamma_-, \, \alpha_{++-}=2\gamma_+-\gamma_-$ and the three point function $\langle \widetilde{\mathcal{O}'}_i \widetilde{\mathcal{O}'}_j \widetilde{\mathcal{O}'}_k\rangle $ is  necessarily conformal up to linear order in $\epsilon$.

\section{Thermal physics}
\label{sec: thermal physics}

To understand the unbroken symmetries of the critical model at finite temperature we consider the effective potential, $V_\mt{eff}$.  To leading order in $\epsilon_i$, thermal fluctuations simply induce quadratic terms in addition to the quartic potential  \reef{Sdeform}.
Starting from the thermal correlation function for  the generalized free fields we find 
\be
 \langle \phi^a_j(\tau, \vec x) \phi^c_i(0) \rangle_\beta = \sum_{m=-\infty}^\infty {\delta^{ac} \delta_{ij}\over \[ (\tau + m \beta)^2 + \vec x^2\]^{\Delta_{\phi_i}}} 
 \quad \Rightarrow \quad \langle \phi_i^2\rangle_\beta= N x_i {2\,  \zeta(2\Delta_{\phi_i})\over \beta^{2\Delta_{\phi_i}}} ~.
 \label{Tvev}
 \ee
Together with the interaction terms in \reef{Sdeform} this leads to the following effective potential for the zero mode
\be
 V_\mt{eff}(\phi_1,\phi_2; \beta)= \mathcal{M}_{\phi_1}(\beta) \phi_1^2 + \mathcal{M}_{\phi_2}(\beta) \phi_2^2 
 + {g_1 \mu^{\epsilon_1} \over N} (\phi_1^2)^2  + {g_2 \mu^{\epsilon_2} \over N} (\phi_2^2)^2
 + {g_3 \mu^{\epsilon_3} \over N} \phi_1^2 \phi_2^2    ~,
 \label{Veff}
\ee
where  we dropped terms suppressed by the higher powers of $\epsilon$ , and 
\bea
  \mathcal{M}_{\phi_1}(\beta) &=& 2 {g_1\mu^{\epsilon_1} \over N} \Big(1 +{2\over N x_1}\Big) \langle \phi_1^2\rangle_\beta
  + {g_3\mu^{\epsilon_3}\over N} \langle \phi_2^2\rangle_\beta ~,
   \label{M} \\
 \mathcal{M}_{\phi_2}(\beta) &=& 2 {g_2\mu^{\epsilon_2} \over N} \Big(1 +{2\over N x_2}\Big) \langle \phi_2^2\rangle_\beta
  + {g_3\mu^{\epsilon_3}\over N} \langle \phi_1^2\rangle_\beta.
\eea

\noindent
In the absence of quadratic terms\footnote{Note that $\mathcal{M}_{\phi_i}$ has the non-canonical scaling dimension of $\frac{d+\epsilon_{i}}{2}$.}, $\mathcal{M}_{\phi_1}=\mathcal{M}_{\phi_2}=0$, the potential reaches its minimum value at $\phi^2_i=0$. Hence, the system exhibits  full $O(m)\times\mathcal{O}(N-m)$ symmetry at zero temperature. However, finite temperature effects may break the symmetry  provided that $ \mathcal{M}_{\phi_i}<0$. If that occurs, the higher order perturbative corrections cannot restore the symmetry, because multiloop quadratic terms are suppressed by additional powers of  $\epsilon_i$, whereas terms with higher powers of  fields will be subdominant in the vicinity of the origin $\phi^2_i=0$. Therefore, to prove that the symmetry is broken at finite temperature, it is enough to show that the model admits a fixed point where one of the quadratic terms  in the effective potential \eqref{Veff} becomes negative. 

\noindent
Thermal expectation values $\langle \phi_i^2 \rangle_\beta$, given by \reef{Tvev}, are positive, and therefore $\mathcal{M}_{\phi_i}$ can only become negative if some of the critical couplings $g_i$ are negative. Since the couplings $g_1$ and $g_2$ must be positive to ensure stability of the model, the only scenario would be $g_3<0$, while $4g_1g_2\geq g_3^2$ to exclude the runaway behavior. In fact, the potential is always bounded from below as long as the fixed point equations \reef{beta-fn} are satisfied \cite{Rychkov:2018vya,Chai:2020zgq}. We verified explicitly (numerically or analytically) that the stability condition $4g_1g_2\geq g_3^2$ holds in all the examples  discussed below.

\noindent
We first analyze the $N\rightarrow \infty$ limit with $x_1=m/N$  kept fixed.  The effective potential in this case assumes  the following form\footnote{As we explained in section \ref{sec:model},  in the large rank limit the couplings must satisfy $\alpha\to 1$, and therefore $\epsilon_1=\epsilon_2=\epsilon$.}
\bea
 \label{eff_pot}
  V_\text{eff}(z; \beta)= 2 {\mu^\epsilon \over N} \Big(     \langle z \rangle_\beta \, z +{z^2\over 2} \Big) ~,
  \quad z=\sqrt{g_1} \phi_1^2 \pm \sqrt{g_2} \phi_2^2 ~.
\eea
Here we used the leading order relation between the critical couplings, $g_3=\pm2\sqrt{g_1 g_2}$, which follows from \reef{branches}. For positive $g_3$, $\langle z\rangle_\beta>0$, and  the effective potential is minimized by $\phi^2_i=0$  because $\phi_i^2$ cannot be negative. For negative $g_3$ the situation is more nuanced. Provided constraints $\phi_i^2>0$ are satisfied,  the minimum is reached at  
\be
 z=\sqrt{g_1} \phi_1^2 - \sqrt{g_2} \phi_2^2= -  \langle z \rangle_\beta=(x_2 \sqrt{g_2}-x_1 \sqrt{g_1}){2\zeta (d/2)\over \beta^{d/2}} ~, \quad {\rm for}\quad g_3<0 ~,
 \label{moduli}
\ee
where in the expression for $\langle z \rangle_\beta$ we dropped $\epsilon_i$-suppressed terms. 
This defines a one-dimensional family of minima in the space of fields -- a hyperbola in the $(\phi_1, \phi_2)$ plane. 
In the special case when $x_1\sqrt{g_1}=x_2\sqrt{g_2}$,  the parameter $\langle z \rangle_\beta=0$, 
and there is trivial solution $\phi_1^2=\phi_2^2=0$, together with the non-trivial ones $\phi_1^2=\phi_2^2>0$. Hence in this case we can not establish symmetry breaking unless $1/N$ corrections are taken into account. But in all other cases \eqref{moduli}, together with the constraints  $\phi_i^2>0$, necessarily yields the solutions with at least one or both fields being non-zero. 

\begin{figure}[]
\begin{center}
\includegraphics[width=300pt]{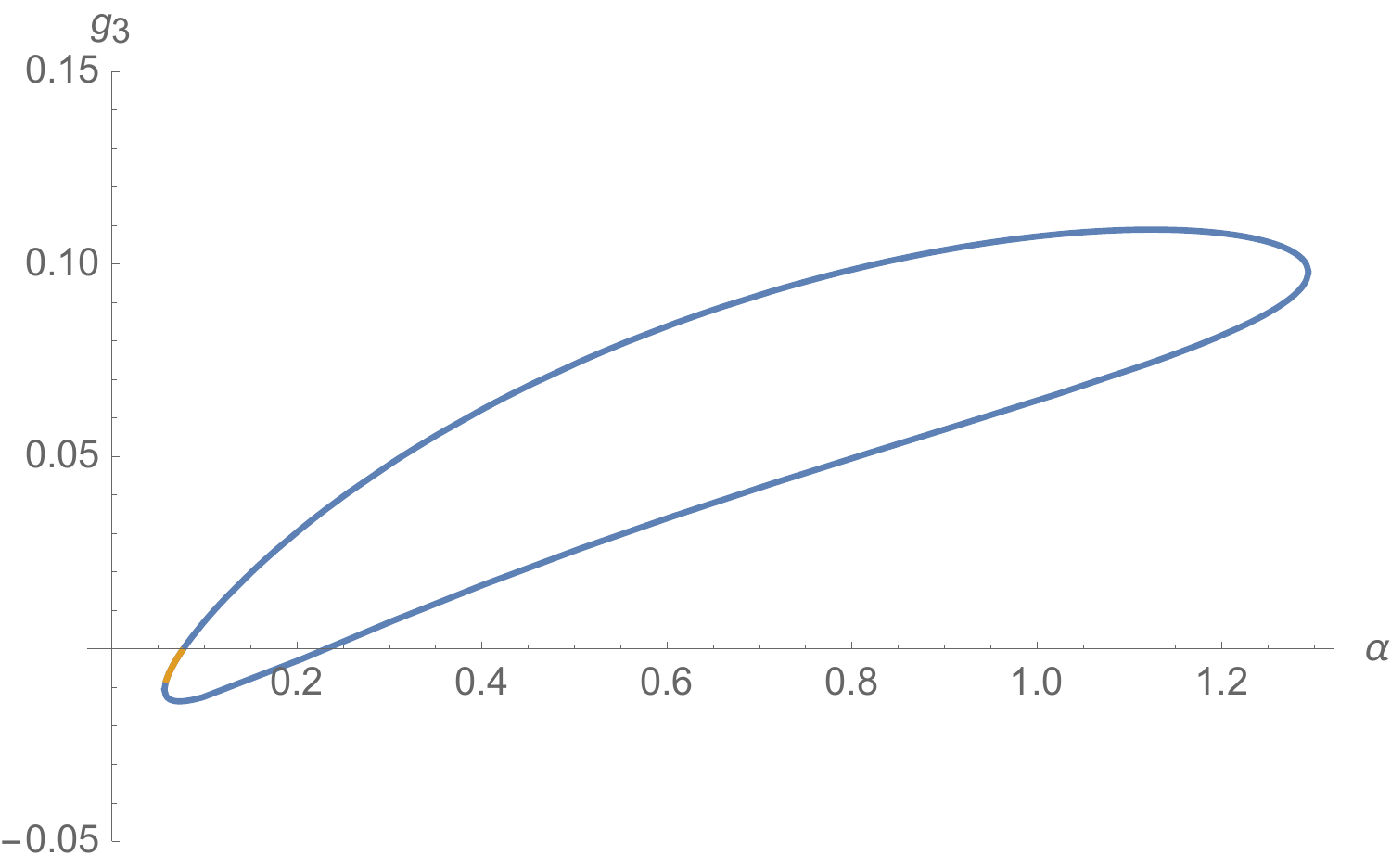}
\end{center}
\caption{Critical value of $g_3$ as a function of $\alpha$ for $m=1,N=6$ case. Orange region corresponds to $\mathcal{M}_{\phi_1}<0$.}
\label{fig:m1N6}
\end{figure}

When $N$ is finite, one in principle needs to minimize \eqref{Veff} with the additional conditions $\phi_i^2>0$. Provided one of the masses is negative, say $\mathcal{M}_{\phi_1}<0$, the minimum is given by 
\bea
\left(\begin{array}{c}\phi_1^2 \\ \phi_2^2 \end{array}\right) =    {-N\mu^{-\epsilon_1} \over 2 g_1}   \left(\begin{array}{c}   \mathcal{M}_{\phi_1} \\0  \end{array}\right) ~.
\eea
while the only point with the unbroken symmetry $\phi_1^2=\phi_2^2=0$ is not  a minimum. It is thus sufficient to show there are critical points with negative $\mathcal{M}_{\phi_1}$, which we do numerically. Thus typically there are two  solutions for critical  $\tilde{g}_i$ for a given value of $\alpha$.  There are points with  $g_3<0$ for which  $\mathcal{M}_{\phi_1}$ is positive, but there are also those where $\mathcal{M}_{\phi_1}<0$. We illustrate that in Fig.~\ref{fig:m1N6} for $m=1,N=6$. 
The behavior for $m=1$ and $N> 6$ as well for $m=2,N\geq 7$, and more generally  $N\geq \min(m+5,2m)$, is similar (we use the notations $m\leq N-m$). There are always values of $\alpha\leq \alpha_c$ such that $g_3,\mathcal{M}_{\phi_1}<0$, and hence the corresponding critical points exhibit persistent symmetry breaking.

\section{Discussion}
\label{sec:discussion}
In this paper we constructed and studied a three-dimensional model comprised of two copies of the generalized free fields $\phi_{1,2}$ in the fundamental representation of  $O(m)$ and $O(N-m)$ respectively.  The bare dimensions  are tuned such that quartic interactions are weakly relevant and consequently the model can be  analyzed using conformal perturbation theory.
One key property of our model is that it exhibits global continuous symmetry breaking at arbitrarily large temperatures. To our knowledge this is the first example of a  UV complete unitary 3d model exhibiting persistent breaking of a continuous global symmetry. It  bypasses the Coleman-Hohenberg-Mermin-Wagner no-go theorem \cite{Mermin:1966fe,Hohenberg:1967zz,Coleman:1973ci}
by incorporating non-local interactions.\footnote{Placing a theory on a curved spacetime is another way to bypass the CHMW theorem. For instance, the O(N) model in AdS evades it \cite{Carmi:2018qzm}, but at high temperatures the symmetry is restored in this model.}

Besides $N$ and $m$ the model is parametrized by $\alpha=\epsilon_1/\epsilon_2$, where $\Delta_{1,2} = (d-\epsilon_{1,2})/4$ are bare scaling dimensions of fundamental fields and we work in the regime $\epsilon_i\ll 1$ to leading order in $\epsilon_i$.

We found that the resulting IR flow terminates at the fixed points located in the perturbative vicinity of the origin. In this sense our model is similar to the  Banks-Zaks  construction \cite{Banks:1981nn}. In the infinite $N$ limit, we find a conformal manifold, which is in fact a circle. It includes points where the coupling constant $g_3$, that controls the coupling  between  $O(m)$ and $O(N-m)$ fields, vanishes, and the model degenerates into two decoupled long range models, one of which is free and the other critical. It also includes a point where global symmetry is enhanced to $O(N)$. For us of particular interest are the fixed points with $g_3<0$, as this is a necessary (but not  sufficient) condition for the
symmetry to be broken at finite temperature.  For large but finite $N$
we have a continuous family of CFTs
parametrized by $\epsilon_i$. The perturbative fixed points with $g_3< 0$ also survive for finite $N$. This behavior continues up to small values of  $N$. Assuming $m\leq N-m$, we find that  $g_3< 0$ fixed points exist for any $N\geq \max(m+5,2m)$.

At zero temperature, $\langle \phi^2_i\rangle=0$ and the full $O(m)\times O(N-m)$ symmetry is preserved. We argue this is a unitary theory with full conformal symmetry, not merely a set of scale-invariant fixed points. For that purpose we show that the anomalous dimensions are real. We also study  two- and three-point functions. 
Working at the linear order in $\epsilon_{1,2}$ 
we demonstrate that  two-point functions of the operators
with different scaling dimensions vanish, as required by full conformal symmetry. Furthermore, within the same approximation three-point functions
exhibit the form consistent with the full conformal invariance. While our results do not constitute a proof, they strongly suggest the interacting $O(m)\times O(N-m)$ theory is unitary and conformal, which extends previous results \cite{Paulos:2015jfa} arguing for conformality of the interacting long range $O(m)$ model. 

At finite temperature $T$ certain fixed points with $g_3<0$, which appear for all $N\geq \max(m+5,2m)$ and particular $\alpha$, exhibit spontaneous symmetry breaking 
$O(m)\times O(N-m) \rightarrow O(m-1)\times O(N-m)$. Since the theory is scale-invariant, symmetry breaking persists at all temperatures. Thus our model provides a generalization of \cite{Chai:2021djc} which considered the case of $m=1$ and reported persistent breaking of discrete $\Z_2=O(1)$ symmetry. The crucial ingredient in the present construction is the non-local interaction, which is necessary to circumvent the CHMW no-go theorem.
In contrast, existing examples of {\it local} theories which exhibit persistent symmetry breaking are either UV-incomplete \cite{Weinberg:1974hy},  or require fractional dimensions \cite{Chai:2020zgq,Chai:2020onq} where unitarity is violated \cite{Hogervorst:2015akt}, or require strictly infinite $N$ \cite{Chaudhuri:2020xxb,Chaudhuri:2021dsq}. In particular, it is still an open question whether persistent breaking of a discrete symmetry is possible in a {\it local}, UV-complete, unitary,  relativistic three-dimensional theory. A candidate for such a model was suggested in \cite{Chai:2020zgq,Chai:2020onq}.\footnote{The example in \cite{Chai:2021djc} is manifestly non-local.} However, it is hard to establish the existence of PSB phenomenon in their model directly in $2+1$ dimensions. A similar question regarding continuous symmetry in $2+1$ dimensions is answered by the CHMW theorem which prohibits such a behavior. It would therefore be interesting to generalize the CHMW no-go  result in the context of {\it local}, UV-complete, unitary,  relativistic three-dimensional theories with discrete symmetries.

Our results prompt further research. Recently, generalized free fermionic models, perturbed by four-fermionic interaction, have been considered in \cite{Chai:2021wac}.
It would be interesting to see if such non-local models lead to conformal fixed points which can exhibit persistent symmetry breaking. Beyond the fermionic QFTs one can ask a similar question already for the lattice settings. It is well-known that in local, i.e.~short range lattice models symmetries are always restored at sufficiently large temperatures, of the order of the inverse lattice size \cite{Bratteli:1979tw}. This is also the case for models with exponentially decaying interactions, which should be regarded as local in all physical senses. The question our work poses is to see if for lattices with the polynomial interactions
the results of \cite{Bratteli:1979tw} break down and if persistent breaking is possible.

{\bf Acknowledgements}  We thank A.~Avdoshkin, S.~Chaudhuri, C.~Choi, Z.~Komargodski, E.~Rabinovici for helpful discussions and correspondence. 
NC, RS and MS are grateful to the Israeli Science Foundation Center of Excellence (grant No. 2289/18) and the Quantum Universe I-CORE program of the Israel Planning and Budgeting Committee (grant No. 1937/12) for continuous support of our research. NC is grateful for the support from the Yuri Milner scholarship.
AD  is supported by the NSF under grant PHY-2013812.
AD also acknowledges KITP for hospitality. The research at KITP was supported in part by the National Science Foundation under Grant No. PHY-1748958. 
The work of MG is partially supported by DOE grant DE-SC0011842.


\newpage

\begin{thebibliography}{99}

\bibitem{Weinberg:1974hy}
S.~Weinberg,
``Gauge and Global Symmetries at High Temperature,''
Phys. Rev. D \textbf{9}, 3357-3378 (1974)
doi:10.1103/PhysRevD.9.3357

\bibitem{Dvali:1995cj}
G.~R.~Dvali, A.~Melfo and G.~Senjanovic,
``Is There a monopole problem?,''
Phys. Rev. Lett. \textbf{75}, 4559-4562 (1995)
doi:10.1103/PhysRevLett.75.4559
[arXiv:hep-ph/9507230 [hep-ph]].

\bibitem{Dvali:1995cc}
G.~R.~Dvali and G.~Senjanovic,
``Is there a domain wall problem?,''
Phys. Rev. Lett. \textbf{74}, 5178-5181 (1995)
doi:10.1103/PhysRevLett.74.5178
[arXiv:hep-ph/9501387 [hep-ph]].

\bibitem{Senjanovic:1998xc}
G.~Senjanovic,
``Rochelle salt: A Prototype of particle physics,''
doi:10.1142/9789814447263\_0062
[arXiv:hep-ph/9805361 [hep-ph]].

\bibitem{Bajc:1999cn}
B.~Bajc,
``High temperature symmetry nonrestoration,''
doi:10.1142/9789812792129\_0039
[arXiv:hep-ph/0002187 [hep-ph]].

\bibitem{Ramazanov:2021eya}
S.~Ramazanov, E.~Babichev, D.~Gorbunov and A.~Vikman,
``Beyond freeze-in: Dark Matter via inverse phase transition and gravitational wave signal,''
[arXiv:2104.13722 [hep-ph]].

\bibitem{Ramazanov:2020ajq}
S.~Ramazanov, F.~R.~Urban and A.~Vikman,
``Observing primordial magnetic fields through Dark Matter,''
JCAP \textbf{02}, 011 (2021)
doi:10.1088/1475-7516/2021/02/011
[arXiv:2010.03383 [astro-ph.CO]].

\bibitem{Chai:2021djc}
N.~Chai, A.~Dymarsky and M.~Smolkin,
``A model of persistent breaking of discrete symmetry,''
[arXiv:2106.09723 [hep-th]].

\bibitem{Hong:2000rk}
S.~I.~Hong and J.~B.~Kogut,
``Symmetry nonrestoration in a Gross-Neveu model with random chemical potential,''
Phys. Rev. D \textbf{63}, 085014 (2001)
doi:10.1103/PhysRevD.63.085014
[arXiv:hep-th/0007216 [hep-th]].

\bibitem{Komargodski:2017dmc}
Z.~Komargodski, A.~Sharon, R.~Thorngren and X.~Zhou,
``Comments on Abelian Higgs Models and Persistent Order,''
SciPost Phys. \textbf{6}, no.1, 003 (2019)
doi:10.21468/SciPostPhys.6.1.003
[arXiv:1705.04786 [hep-th]].

\bibitem{Chai:2020onq}
N.~Chai, S.~Chaudhuri, C.~Choi, Z.~Komargodski, E.~Rabinovici and M.~Smolkin,
``Symmetry Breaking at All Temperatures,''
Phys. Rev. Lett. \textbf{125}, no.13, 131603 (2020)
doi:10.1103/PhysRevLett.125.131603

\bibitem{Chai:2020zgq}
N.~Chai, S.~Chaudhuri, C.~Choi, Z.~Komargodski, E.~Rabinovici and M.~Smolkin,
``Thermal Order in Conformal Theories,''
Phys. Rev. D \textbf{102}, no.6, 065014 (2020)
doi:10.1103/PhysRevD.102.065014
[arXiv:2005.03676 [hep-th]].

\bibitem{Chai:2020hnu}
N.~Chai, E.~Rabinovici, R.~Sinha and M.~Smolkin,
``The bi-conical vector model at $1/N$,''
JHEP \textbf{05}, 192 (2021)
doi:10.1007/JHEP05(2021)192
[arXiv:2011.06003 [hep-th]].

\bibitem{Chaudhuri:2020xxb}
S.~Chaudhuri, C.~Choi and E.~Rabinovici,
``Thermal order in large N conformal gauge theories,''
JHEP \textbf{04}, 203 (2021)
doi:10.1007/JHEP04(2021)203
[arXiv:2011.13981 [hep-th]].

\bibitem{Bajc:2020gpa}
B.~Bajc, A.~Lugo and F.~Sannino,
``Asymptotically free and safe fate of symmetry nonrestoration,''
Phys. Rev. D \textbf{103}, 096014 (2021)
doi:10.1103/PhysRevD.103.096014
[arXiv:2012.08428 [hep-th]].

\bibitem{Chaudhuri:2021dsq}
S.~Chaudhuri and E.~Rabinovici,
``Symmetry breaking at high temperatures in large N gauge theories,''
JHEP \textbf{08}, 148 (2021)
doi:10.1007/JHEP08(2021)148
[arXiv:2106.11323 [hep-th]].

\bibitem{Buchel:2009ge}
A.~Buchel and C.~Pagnutti, ``{Exotic Hairy Black Holes},''
 {\em Nucl. Phys. B}
  {\bf 824} (2010)  85--94,
  [arXiv:0904.1716 [hep-th]].

\bibitem{Donos:2011ut}
A.~Donos and J.~P. Gauntlett, ``{Superfluid black branes in $AdS_4\times
  S^7$},''{\em JHEP} {\bf 06}
  (2011)  053, 
  [arXiv:1104.4478[hep-th]].

\bibitem{Gursoy:2018umf}
U.~G\"ursoy, E.~Kiritsis, F.~Nitti, and L.~Silva~Pimenta, ``{Exotic holographic
  RG flows at finite temperature},''
  {\em JHEP} {\bf 10} (2018)
  173,
  [arXiv:1805.01769[hep-th]].

\bibitem{Buchel:2018bzp}
A.~Buchel, ``{Klebanov-Strassler black hole},''
 {\em JHEP} {\bf 01} (2019)
  207,
  [arXiv:1809.08484[hep-th]].

\bibitem{Buchel:2020thm}
A.~Buchel, ``{Thermal order in holographic CFTs and no-hair theorem violation
  in black branes},''
 {\em Nucl. Phys. B}
  {\bf 967} (2021)  115425,
  [arXiv:2005.07833 [hep-th]].

\bibitem{Buchel:2020xdk}
A.~Buchel, ``{Holographic conformal order in supergravity},''
 {\em Phys. Lett. B}
  {\bf 814} (2021)  136111,
  [arXiv:2007.09420 [hep-th]].

\bibitem{Buchel:2020jfs}
A.~Buchel, ``{Fate of the conformal order},''
 {\em Phys. Rev. D} {\bf
  103} (2021) no.~2, 026008,
  [arXiv:2011.11509 [hep-th]].
  

\bibitem{Buchel:2021ead}
A.~Buchel,
``Compactified holographic conformal order,''
[arXiv:2107.05086 [hep-th]].

\bibitem{Tanizaki:2017qhf}
Y.~Tanizaki, T.~Misumi and N.~Sakai,
``Circle compactification and \textquoteright{}t Hooft anomaly,''
JHEP \textbf{12}, 056 (2017)
doi:10.1007/JHEP12(2017)056
[arXiv:1710.08923 [hep-th]].

\bibitem{Dunne:2018hog}
G.~V.~Dunne, Y.~Tanizaki and M.~\"Unsal,
``Quantum Distillation of Hilbert Spaces, Semi-classics and Anomaly Matching,''
JHEP \textbf{08}, 068 (2018)
doi:10.1007/JHEP08(2018)068
[arXiv:1803.02430 [hep-th]].

\bibitem{Wan:2019oax}
Z.~Wan and J.~Wang,
``Higher anomalies, higher symmetries, and cobordisms III: QCD matter phases anew,''
Nucl. Phys. B \textbf{957}, 115016 (2020)
doi:10.1016/j.nuclphysb.2020.115016
[arXiv:1912.13514 [hep-th]].

\bibitem{Aitken:2017ayq}
K.~Aitken, A.~Cherman, E.~Poppitz and L.~G.~Yaffe,
``QCD on a small circle,''
Phys. Rev. D \textbf{96}, no.9, 096022 (2017)
doi:10.1103/PhysRevD.96.096022
[arXiv:1707.08971 [hep-th]].

\bibitem{Mermin:1966fe}
N.~D.~Mermin and H.~Wagner,
``Absence of ferromagnetism or antiferromagnetism in one-dimensional or two-dimensional isotropic Heisenberg models,''
Phys. Rev. Lett. \textbf{17}, 1133-1136 (1966)
doi:10.1103/PhysRevLett.17.1133

\bibitem{Hohenberg:1967zz}
P.~C.~Hohenberg,
``Existence of Long-Range Order in One and Two Dimensions,''
Phys. Rev. \textbf{158}, 383-386 (1967)
doi:10.1103/PhysRev.158.383

\bibitem{Coleman:1973ci}
S.~R.~Coleman,
``There are no Goldstone bosons in two-dimensions,''
Commun. Math. Phys. \textbf{31}, 259-264 (1973)
doi:10.1007/BF01646487

\bibitem{Mermin:1968zz}
N.~D.~Mermin,
``Crystalline Order in Two Dimensions,''
Phys. Rev. \textbf{176}, 250-254 (1968)
doi:10.1103/PhysRev.176.250

\bibitem{Halperin_2018}
B.~I. Halperin, ``On the
  Hohenberg-Mermin-Wagner theorem and its limitations,''{\em Journal of
  Statistical Physics} {\bf 175} (Dec, 2018)  521-529.
  http://dx.doi.org/10.1007/s10955-018-2202-y.

\bibitem{Hogervorst:2015akt}
M.~Hogervorst, S.~Rychkov and B.~C.~van Rees,
``Unitarity violation at the Wilson-Fisher fixed point in 4-$\epsilon$ dimensions,''
Phys. Rev. D \textbf{93}, no.12, 125025 (2016)
doi:10.1103/PhysRevD.93.125025
[arXiv:1512.00013 [hep-th]].

\bibitem{Fisher:1972zz}
M.~E.~Fisher, S.~k.~Ma and B.~G.~Nickel,
``Critical Exponents for Long-Range Interactions,''
Phys. Rev. Lett. \textbf{29}, 917-920 (1972)
doi:10.1103/PhysRevLett.29.917

\bibitem{Wilson:1971dc}
K.~G.~Wilson and M.~E.~Fisher,
``Critical exponents in 3.99 dimensions,''
Phys. Rev. Lett. \textbf{28}, 240-243 (1972)
doi:10.1103/PhysRevLett.28.240

\bibitem{Paulos:2015jfa}
M.~F.~Paulos, S.~Rychkov, B.~C.~van Rees and B.~Zan,
``Conformal Invariance in the Long-Range Ising Model,''
Nucl. Phys. B \textbf{902}, 246-291 (2016)
doi:10.1016/j.nuclphysb.2015.10.018
[arXiv:1509.00008 [hep-th]].

\bibitem{Behan:2017dwr}
C.~Behan, L.~Rastelli, S.~Rychkov and B.~Zan,
``Long-range critical exponents near the short-range crossover,''
Phys. Rev. Lett. \textbf{118}, no.24, 241601 (2017)
doi:10.1103/PhysRevLett.118.241601
[arXiv:1703.03430 [cond-mat.stat-mech]].

\bibitem{Behan:2017emf}
C.~Behan, L.~Rastelli, S.~Rychkov and B.~Zan,
``A scaling theory for the long-range to short-range crossover and an infrared duality,''
J. Phys. A \textbf{50}, no.35, 354002 (2017)
doi:10.1088/1751-8121/aa8099
[arXiv:1703.05325 [hep-th]].

\bibitem{Behan:2018hfx}
C.~Behan,
``Bootstrapping the long-range Ising model in three dimensions,''
J. Phys. A \textbf{52}, no.7, 075401 (2019)
doi:10.1088/1751-8121/aafd1b
[arXiv:1810.07199 [hep-th]].

\bibitem{Gubser:2017vgc}
S.~S.~Gubser, C.~Jepsen, S.~Parikh and B.~Trundy,
``O(N) and O(N) and O(N),''
JHEP \textbf{11}, 107 (2017)
doi:10.1007/JHEP11(2017)107
[arXiv:1703.04202 [hep-th]].

\bibitem{Giombi:2019enr}
S.~Giombi and H.~Khanchandani,
``$O(N)$ models with boundary interactions and their long range generalizations,''
JHEP \textbf{08}, no.08, 010 (2020)
doi:10.1007/JHEP08(2020)010
[arXiv:1912.08169 [hep-th]].

\bibitem{Chai:2021arp}
N.~Chai, M.~Goykhman and R.~Sinha,
``Long-Range Vector Models at Large N,''
[arXiv:2107.08052 [hep-th]].

\bibitem{Chakraborty:2021lwl}
S.~Chakraborty and M.~Goykhman,
``Critical long-range vector model in the UV,''
[arXiv:2108.10084 [hep-th]].

\bibitem{Brydges:2002wq}
D.~C.~Brydges, P.~K.~Mitter and B.~Scoppola,
``Critical (Phi**4)(3, epsilon),''
Commun. Math. Phys. \textbf{240}, 281-327 (2003)
doi:10.1007/s00220-003-0895-4
[arXiv:hep-th/0206040 [hep-th]].

\bibitem{Abdesselam:2006qg}
A.~Abdesselam,
``A Complete Renormalization Group Trajectory Between Two Fixed Points,''
Commun. Math. Phys. \textbf{276}, 727-772 (2007)
doi:10.1007/s00220-007-0352-x
[arXiv:math-ph/0610018 [math-ph]].

\bibitem{Brezin2014}
E.~Brezin, G.~Parisi, and F.~Ricci-Tersengh,
``The Crossover Region Between
  Long-Range and Short-Range Interactions for the Critical Exponents,''
Journal of
  Statistical Physics, no.4-5, 010 (2014)
doi:10.1007/s10955-014-1081-0
[arXiv:1407.3358v1 [cond-mat.stat-mech]].

\bibitem{Slade:2016yer}
G.~Slade,
``Critical Exponents for Long-Range ${O(n)}$ Models Below the Upper Critical Dimension,''
Commun. Math. Phys. \textbf{358}, no.1, 343-436 (2018)
doi:10.1007/s00220-017-3024-5
[arXiv:1611.06169 [math-ph]].

\bibitem{Benedetti:2020rrq}
D.~Benedetti, R.~Gurau, S.~Harribey and K.~Suzuki,
``Long-range multi-scalar models at three loops,''
J. Phys. A \textbf{53}, no.44, 445008 (2020)
doi:10.1088/1751-8121/abb6ae
[arXiv:2007.04603 [hep-th]].

\bibitem{Rychkov:2018vya}
S.~Rychkov and A.~Stergiou,
``General Properties of Multiscalar RG Flows in $d=4-\varepsilon$,''
SciPost Phys. \textbf{6}, no.1, 008 (2019)
doi:10.21468/SciPostPhys.6.1.008
[arXiv:1810.10541 [hep-th]].

\bibitem{Carmi:2018qzm}
D.~Carmi, L.~Di Pietro and S.~Komatsu,
``A Study of Quantum Field Theories in AdS at Finite Coupling,''
JHEP \textbf{01}, 200 (2019)
doi:10.1007/JHEP01(2019)200
[arXiv:1810.04185 [hep-th]].

\bibitem{Banks:1981nn}
T.~Banks and A.~Zaks,
``On the Phase Structure of Vector-Like Gauge Theories with Massless Fermions,''
Nucl. Phys. B \textbf{196}, 189-204 (1982)
doi:10.1016/0550-3213(82)90035-9

\bibitem{Chai:2021wac}
N.~Chai, S.~Chakraborty, M.~Goykhman and R.~Sinha,
``Long-range fermions and critical dualities,''
[arXiv:2110.00020 [hep-th]].

\bibitem{Bratteli:1979tw}
O.~Bratteli and D.~W.~Robinson,
``Operator algebras and quantum statistical mechanics. 1. C* and W* algebras,
symmetry groups, decomposition of states,'' Springer-verl.(1979).


\bibitem{Behan:2017emf}
C.~Behan, L.~Rastelli, S.~Rychkov and B.~Zan,
J. Phys. A \textbf{50} (2017) no.35, 354002
doi:10.1088/1751-8121/aa8099
[arXiv:1703.05325 [hep-th]].
 
 
\bibitem{Komargodski:2016auf}
Z.~Komargodski and D.~Simmons-Duffin,
J. Phys. A \textbf{50} (2017) no.15, 154001
doi:10.1088/1751-8121/aa6087
[arXiv:1603.04444 [hep-th]].


\bibitem{Cardy:1996xt}
J.~L.~Cardy,
``Scaling and renormalization in statistical physics'', Cambridge, UK: Univ. Pr. (1996) 238 p. (Cambridge lecture notes in physics: 3)



\bibitem{Cappelli:1991ke}
A.~Cappelli, J.~I.~Latorre and X.~Vilasis-Cardona,
Nucl. Phys. B \textbf{376} (1992), 510-538
doi:10.1016/0550-3213(92)90119-V
[arXiv:hep-th/9109041 [hep-th]].



\end{thebibliography}

\end{document}